\documentclass[12pt]{article}
\usepackage{graphicx}
\usepackage[margin=1.0in]{geometry}
\usepackage{amsmath}
\usepackage{amssymb} 
\usepackage{hyperref}
\usepackage[affil-it]{authblk}
\usepackage[labelsep=quad,indention=10pt]{subfig}
\usepackage{xcolor}

\begin{document}

\title{Letter of Interest for\\a Neutrino Beam from Protvino to KM3NeT/ORCA}


\author[1]{A.~V.~Akindinov}     
\author[2]{E.~G.~Anassontzis}	
\author[3]{G.~Anton}		
\author[4]{M.~Ardid}		
\author[5]{J.~Aublin}		
\author[5]{B.~Baret}		
\author[6]{V.~Bertin}		
\author[5]{S.~Bourret}		
\author[7]{C.~Bozza}		
\author[3]{M.~Bruchner}		
\author[8,9]{R.~Bruijn}		
\author[6]{J.~Brunner}		
\author[10]{M.~Chabab}		
\author[5]{N.~Chau}		
\author[11]{A.~S.~Chepurnov}	
\author[5,12]{M.~Colomer~Molla}	
\author[6]{P.~Coyle}		
\author[5]{A.~Creusot}		
\author[5]{G.~de~Wasseige}	
\author[6,13,14]{A.~Domi}	
\author[5]{C.~Donzaud}		
\author[3]{T.~Eberl}		
\author[3,6]{A.~Enzenh\"ofer}	
\author[15]{M.~Faifman}		
\author[16]{M.~D.~Filipovi\'c}	
\author[5]{L.~Fusco}		
\author[17]{V.~I.~Garkusha}	
\author[3]{T.~Gal}		
\author[12]{S.~R.~Gozzini}	
\author[3]{K.~Graf}		
\author[5]{T.~Gr\'egoire}	
\author[7]{G.~Grella}		
\author[3]{S.~Hallmann}		
\author[8]{A.~Heijboer}		
\author[12]{J.~J.~Hern\'andez-Rey}	
\author[3]{J.~Hofest\"adt}	
\author[17]{S.~V.~Ivanov}	
\author[18]{C.~W.~James}	
\author[8]{M.~de~Jong}		
\author[8,9]{P.~de~Jong}	
\author[19]{P.~Kalaczy\'nski}	
\author[20]{I.~D.~Kakorin}	
\author[3]{U.~F.~Katz}		
\author[12]{N.~R.~Khan~Chowdhury}	
\author[21]{M.~M.~Kirsanov}	
\author[5]{A.~Kouchner}	
\author[13]{V.~Kulikovskiy}	
\author[1,15,20]{K.~S.~Kuzmin}	
\author[5]{R.~Le Breton}	
\author[17]{O.~P.~Lebedev}	
\author[6]{M.~Lincetto}		
\author[15,22]{E.~Litvinovich}	
\author[23]{D.~Lopez-Coto}	
\author[24]{C.~Markou}		
\author[17]{A.~V.~Maximov}	
\author[8]{K.~W.~Melis}		
\author[8]{R.~Muller}		
\author[20]{V.~A.~Naumov}	
\author[23]{S.~Navas}		
\author[8]{L.~Nauta}		
\author[5]{C.~Nielsen}		
\author[17]{F.~N.~Novoskoltsev}	
\author[8,9]{B.~\'O~Fearraigh}	
\author[25]{M.~Organokov}	
\author[26]{G.~Papalashvili}	
\author[6]{M.~Perrin-Terrin}	
\author[4]{C.~Poir\`e}		
\author[25]{T.~Pradier} 	
\author[6]{L.~Quinn}		
\author[8]{D.~F.~E.~Samtleben}	
\author[13]{M.~Sanguineti}	
\author[8]{J.~Seneca}		
\author[26]{R.~Shanidze}	
\author[11]{E.~V.~Shirokov}	
\author[24]{A.~Sinopoulou}	
\author[17]{R.~Yu.~Sinyukov}	
\author[15,22]{M.~D.~Skorokhvatov}
\author[15]{I.~Sokalski}	
\author[17]{A.~A.~Sokolov}	
\author[7,27]{B.~Spisso}	
\author[7,27]{S.~M.~Stellacci}	
\author[8]{B.~Strandberg}	
\author[13,14]{M.~Taiuti}	
\author[12]{T.~Thakore}		
\author[24]{E.~Tzamariudaki}	
\author[5]{V.~Van~Elewyck}	
\author[8,9]{E.~de~Wolf}	
\author[1,6]{D.~Zaborov\thanks{Corresponding author. E-mail address: {\textit{zaborov@itep.ru}}.}}	
\author[17]{A.~M.~Zaitsev}	
\author[12]{J.~D.~Zornoza}	
\author[12]{J.~Z\'u\~niga} 	

\affil[1]{A.I. Alikhanov Institute for Theoretical and Experimental Physics of NRC~``Kurchatov~Institute'', Moscow, Russia}
\affil[2]{Physics Department, N. and K. University of Athens, Athens, Greece}
\affil[3]{Friedrich-Alexander-Universit\"at Erlangen-N\"urnberg, Erlangen Centre for Astroparticle Physics, Erlangen, Germany}
\affil[4]{Universitat Polit\`ecnica de Val\`encia, Instituto de Investigaci\'on para la Gesti\'on Integrada de las Zonas Costeras, Gandia, Spain}
\affil[5]{APC, Universit\'e Paris Diderot, CNRS/IN2P3, CEA/IRFU, Observatoire de Paris, Sorbonne~Paris~Cit\'e, Paris, France}
\affil[6]{Aix Marseille Univ, CNRS/IN2P3, CPPM, Marseille, France}
\affil[7]{Universit\`a di Salerno e INFN Gruppo Collegato di Salerno, Dipartimento di Fisica, Fisciano, Italy}
\affil[8]{Nikhef, National Institute for Subatomic Physics, Amsterdam, Netherlands}
\affil[9]{University of Amsterdam, Institute of Physics/IHEF, Amsterdam, Netherlands}
\affil[10]{Cadi Ayyad University, Physics Department, Faculty of Science Semlalia, Marrakech, Morocco}
\affil[11]{D.V. Skobeltsyn Institute of Nuclear Physics, Moscow State University, Moscow, Russia}
\affil[12]{IFIC -- Instituto de F\'isica Corpuscular (CSIC -- Universitat de Val\`encia), Valencia, Spain}
\affil[13]{INFN, Sezione di Genova, Genova, Italy}
\affil[14]{Universit\`a di Genova, Genova, Italy}
\affil[15]{National Research Centre ``Kurchatov Institute'', Moscow, Russia}
\affil[16]{Western Sydney University, School of Computing, Engineering and Mathematics, Penrith, Australia}
\affil[17]{A.A. Logunov Institute for High Energy Physics of NRC ``Kurchatov~Institute'', Protvino, Russia}
\affil[18]{Curtin Institute of Radio Astronomy, Curtin University, Bentley, Australia}
\affil[19]{National Centre for Nuclear Research, Warsaw, Poland}
\affil[20]{Joint Institute for Nuclear Research, Dubna, Russia}
\affil[21]{Institute for Nuclear Research of the Russian Academy of Sciences, Moscow, Russia}
\affil[22]{National Research Nuclear University MEPhI (Moscow Engineering Physics Institute), Moscow,~Russia}
\affil[23]{University of Granada, Dpto. de F\'isica Te\'orica y del Cosmos \& C.A.F.P.E., Granada, Spain}
\affil[24]{Institute of Nuclear and Particle Physics, NCSR Demokritos, Athens, Greece}
\affil[25]{Universit\'e de Strasbourg, CNRS, IPHC, Strasbourg, France}
\affil[26]{Tbilisi State University, Department of Physics, Tbilisi, Georgia}
\affil[27]{INFN, Sezione di Napoli, Complesso Universitario di Monte S. Angelo, Napoli, Italy}

\date{}

\maketitle

\sloppy

\begin{abstract}
The Protvino accelerator facility located in the Moscow region, Russia, is in a good position to offer a rich experimental research program in the field of neutrino physics.
Of particular interest is the possibility to direct a neutrino beam from Protvino towards the KM3NeT/ORCA detector,
which is currently under construction in the Mediterranean Sea 40~km offshore Toulon, France.
This proposal is known as P2O.
Thanks to its baseline of 2595 km, this experiment
would yield an unparalleled sensitivity to matter effects in the Earth,
allowing for the determination of the neutrino mass ordering
with a high level of certainty after only a few years of running
at a modest beam intensity of $\approx$~90~kW.
With a prolonged exposure ($\approx$ 1500 kW\,$\cdot$\,yr),
a 2$\sigma$ sensitivity to the leptonic CP-violating Dirac phase can be achieved.
A second stage of the experiment,
comprising a further intensity upgrade of the accelerator complex
and a densified version of the ORCA detector (Super-ORCA),
would allow for up to a 6$\sigma$ sensitivity to CP violation 
and a 10$^\circ$--17$^\circ$ resolution on the CP phase after 10 years of running with a 450 kW beam,
competitive with other planned experiments.
The initial composition and energy spectrum of the neutrino beam would need to be monitored by a near detector,
to be constructed several hundred meters downstream from the proton beam target.
The same neutrino beam and near detector set-up would also allow for neutrino-nucleus cross section measurements to be performed.
A short-baseline sterile neutrino search experiment would also be possible.
\end{abstract}

\section{Introduction}
Neutrino physics is one of the most actively developing branches of particle physics, with many fundamental parameters still awaiting to be experimentally determined, and shows great promise for new insights into physics beyond the Standard Model.
Two of the key open questions are the presence of charge-parity (CP) violation in the lepton sector,
e.g. by the CP-violating Dirac phase in the neutrino mixing matrix,
and the relative ordering of the three neutrino mass eigenstates (``mass ordering'').
Both questions can be answered by studying flavour oscillations of GeV neutrinos over a long baseline ($\gg$ 100~km).
Particle accelerators provide a well-controlled environment suited for conducting high precision measurements of that type.
Several long-baseline accelerator neutrino experiments are currently running and/or under construction,
in particular the T2K/T2HK experiment in Japan (295~km baseline) \cite{T2K,T2HK_2015}, the NO$\nu$A experiment in the USA (810~km baseline) \cite{NOVA_TDR},
and the DUNE experiment (1300~km baseline), also in the USA \cite{DUNE_CDR,DUNE_IDR}.
A typical set-up includes a near detector, to measure the initial energy spectrum and composition of the neutrino beam, and a far detector, to measure the neutrino beam properties after oscillations.
Several experiments with different baselines will likely be necessary to cleanly disentangle effects from various poorly constrained parameters,
such as the CP-violating phase $\delta_{\textrm{CP}}$, the mass ordering, and (the octant of) the $\theta_{23}$ mixing angle.
Furthermore, any new significant experimental finding will need to be independently verified, ideally with an experiment which does not share the same systematic measurement uncertainties.
In this regard, the construction of multiple experiments with different baselines is generally well motivated.

This letter expresses interest in a long-baseline neutrino experiment using the accelerator complex in Protvino (Moscow Oblast, Russia)
to generate a neutrino beam and using the KM3NeT/ORCA detector \cite{KM3NeT_LoI} in the Mediterranean Sea as a far detector.
The scientific potential of the Protvino-ORCA (P2O) experiment is presented
with an emphasis on the sensitivity to the CP-violating Dirac phase $\delta_{\textrm{CP}}$ and neutrino mass ordering.
We argue that, thanks to the long baseline (2595 km) and the 8~megaton sensitive volume of the far detector,
P2O would be complementary and competitive with experiments such as T2K, NO$\nu$A and DUNE.
A vision of the long-term future of P2O is proposed, including upgrades of the Protvino accelerator complex and the ORCA detector.
Additionally, a short-baseline neutrino research program is proposed which includes studies of neutrino-nucleus interactions
as well as searches for phenomena beyond the Standard Model.

This document is organized as follows:
the ORCA neutrino detector is introduced in Section~\ref{sect:orca}.
The current status and proposed upgrades of the Protvino accelerator complex are presented in Section~\ref{sect:U70}.
The neutrino beamline and the near detector are discussed in Sections~\ref{sect:beamline} and \ref{sect:near_detector}, respectively.
Sections~\ref{sect:science_far_detector} and \ref{sect:science_near_detector} present the scientific potential
of the P2O long-baseline experiment and the proposed short-baseline research program, respectively.
Section~\ref{sect:future} refers to a possible future upgrade of ORCA.
Section~\ref{sect:summary} gives a summary.

\section{KM3NeT/ORCA}
\label{sect:orca}

\begin{figure}
  \centering
  \includegraphics[width=8.5cm]{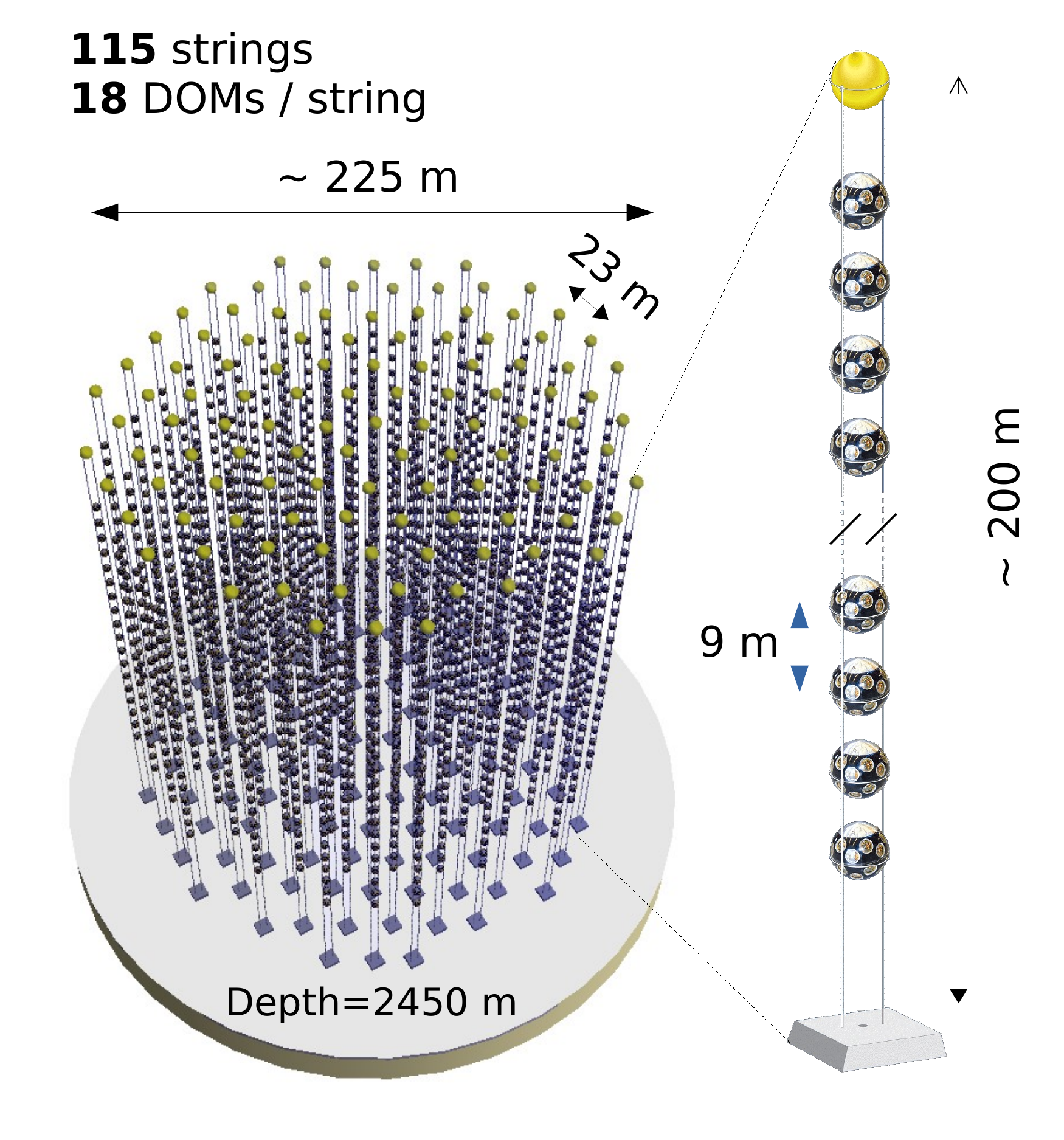}
  \caption{Schematic view of the KM3NeT/ORCA detector.}
  \label{fig:orca_schematic}
\end{figure}

ORCA (Oscillation Research with Cosmics in the Abyss) is one of the two neutrino detectors under construction by the KM3NeT Collaboration \cite{KM3NeT_LoI}.
It is located at 42$^\circ$\,48'\,N 06$^\circ$\,02'\,E, about 40 km off the coast of Toulon, France, at a depth between 2450 m (the seabed depth) and 2250 m.
When completed, ORCA will consist of 2070 digital optical modules (DOMs) installed on 115 vertical strings (detection units, DUs) (see Fig.~\ref{fig:orca_schematic}).
With a 9 m vertical spacing between the DOMs and a $\approx \, 20$ m horizontal spacing between the DUs,
the detector instruments a total of 8 megaton (Mt) of sea water.
ORCA is optimized for the study of atmospheric neutrino oscillations in the energy range of 2 to 30 GeV with the primary goal to determine the neutrino mass ordering.
The majority of neutrino events observed by ORCA will be from electron and muon neutrino and antineutrino charge-current (CC) interactions,
while tau neutrinos and neutral current (NC) interactions constitute minor backgrounds
(7\% and 11\% of the total neutrino rate, respectively, for $\nu_\tau$ CC and all-flavour NC).
At $E_\nu$ = 5~GeV, the majority ($>$~50\%) of muon neutrino CC events detected by ORCA can be correctly identified as muon neutrinos, while less than 15\% of electron neutrino CC events are misidentified as muon neutrinos \cite{KM3NeT_LoI}.
ORCA will provide a neutrino energy resolution of $\approx$~30\% and a zenith angle resolution of $\approx$ 7 degrees at $E_\nu$ = 5~GeV.
A result with a 3$\sigma$ statistical significance for the type of mass ordering is expected after three years of data taking \cite{KM3NeT_LoI}.
ORCA will also provide improved measurements of the atmospheric neutrino oscillation parameters $\Delta m_{23}^2$, $\theta_{23}$
and will probe the unitarity of 3-neutrino mixing by measuring the $\nu_\tau$ flux normalisation.
Non-standard neutrino interactions, as well as astrophysical neutrino sources, dark matter, and other physics phenomena will also be studied. 
The detector construction has recently started and is expected to be completed within 4 years.

\section{The Protvino Accelerator Complex, Current Status and Proposed Upgrades}
\label{sect:U70}

\begin{figure}
  \centering
  \includegraphics[width=12cm]{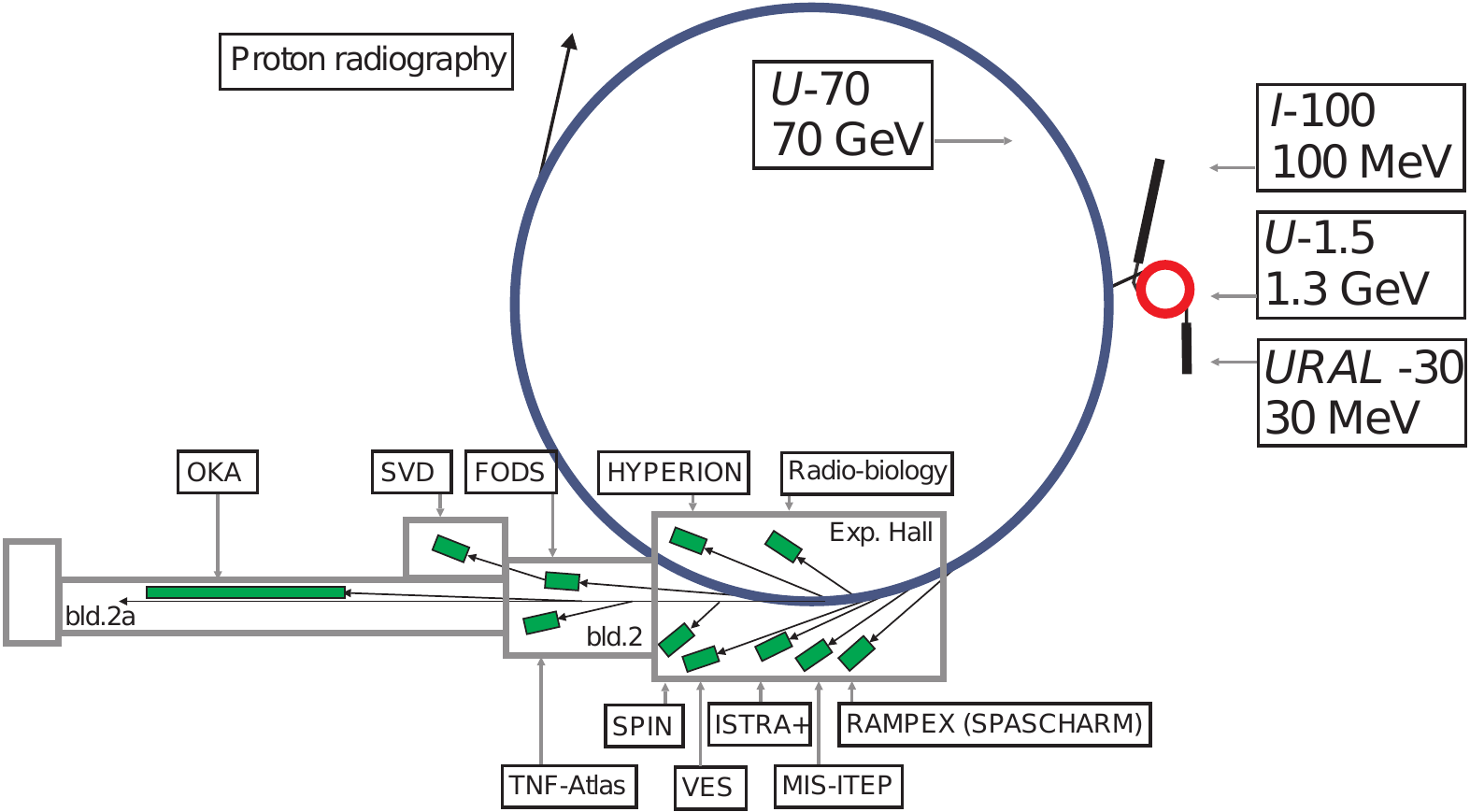}
  \hfill
  \caption{Schematic view of the Protvino accelerator complex.}
  \label{fig:protvino_schematic}
\end{figure}

The Protvino accelerator complex (see Fig.~\ref{fig:protvino_schematic})
is located at 54$^\circ$\,52'\,N 37$^\circ$\,11'\,E, approximately 100 km South of Moscow, Russia.
Its core component is the U-70 synchrotron with a circumference of 1.5 km which accelerates protons up to 70 GeV.
U-70 was originally built in the 1960s and has been in regular operation since then.
The proton injection chain includes an ion source, a 30 MeV linear accelerator, and a 1.5 GeV booster synchrotron.
The accelerator chain is normally operated at a beam energy of 50 GeV to 70 GeV,
with a proton intensity of up to $1.5 \times 10^{13}$ protons per cycle.
The beam cycle is 10 s, with a beam spill duration of up to 3.5 s; or 8 s, with a 5 $\mu$s beam spill.
A dedicated neutrino beamline supplied a neutrino beam to 
the SKAT bubble chamber (1974--1992) \cite{Ammosov1992},
the ITEP-IHEP spark chamber spectrometer \cite{Bugorsky1978},
the IHEP-JINR neutrino detector (1989--1995, upgraded 2002--2006) \cite{Barabash2002},
and other experiments.
The results from these experiments include neutrino-nucleon cross section measurements
and constraints on the $\nu_\mu \rightarrow \nu_e$ oscillation parameters.
The beamline was able to provide a high-purity muon neutrino beam, thanks to the steel muon absorbers preventing muon decay in flight,
and a tunable beam spectrum, thanks to active lenses.
The beamline is not currently operational and its active components will require refurbishing if they are to be used again.
Meanwhile, the rest of the U-70 accelerator complex is in good operational condition.
The complex is operated by the Institute for High Energy Physics (IHEP), which is part of the ``Kurchatov Institute'' National Research Center.

The U-70 synchrotron routinely operates at a time-averaged beam power of up to 15~kW.
In the 1990s, a new injection scheme was considered at IHEP,
which would allow for an increase of the beam intensity to $5 \times 10^{13}$ protons per cycle \cite{UNK_1993}. 
Together with the shortening of the cycle to 7 s, this would provide a beam power of 75 kW.
After some further incremental improvements, a beam power of 90 kW could be reached.
Hence, in the following, we will use the value of 90 kW as the achievable goal of such an upgrade.
Assuming that the accelerator works for the neutrino program with a 60\% efficiency for 6 months a year,
one year of the 90~kW beam corresponds to $\approx 0.8 \times 10^{20}$ protons on target (POT).
Note that the design of the main U-70 synchrotron potentially allows for operation at a beam power up to $\approx$ 450~kW.
An upgrade up to 450~kW could be made possible by a new chain of injection accelerators \cite{OMEGA_LoI}.
Such a beam power would be adequate for high-precision studies of CP violation (see Sect.~\ref{sect:future}).

\section{Neutrino Beamline}
\label{sect:beamline}
\begin{figure}
  \centering
  \includegraphics[width=9.0cm]{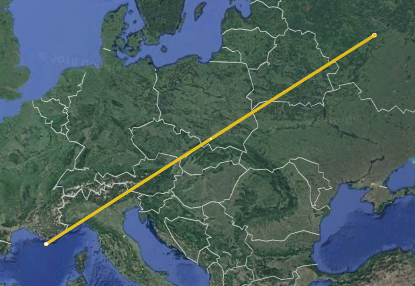}
  \caption{Path to be traveled by the neutrino beam from Protvino (in the top right) to ORCA (in the bottom left).
	   The path length is $\approx$ 2595 km and the deepest point is 135 km below sea level, in the upper mantle.}
  \label{fig:p2o_beam_path}
\end{figure}

\begin{figure}
  \centering
    \includegraphics[width=14cm]{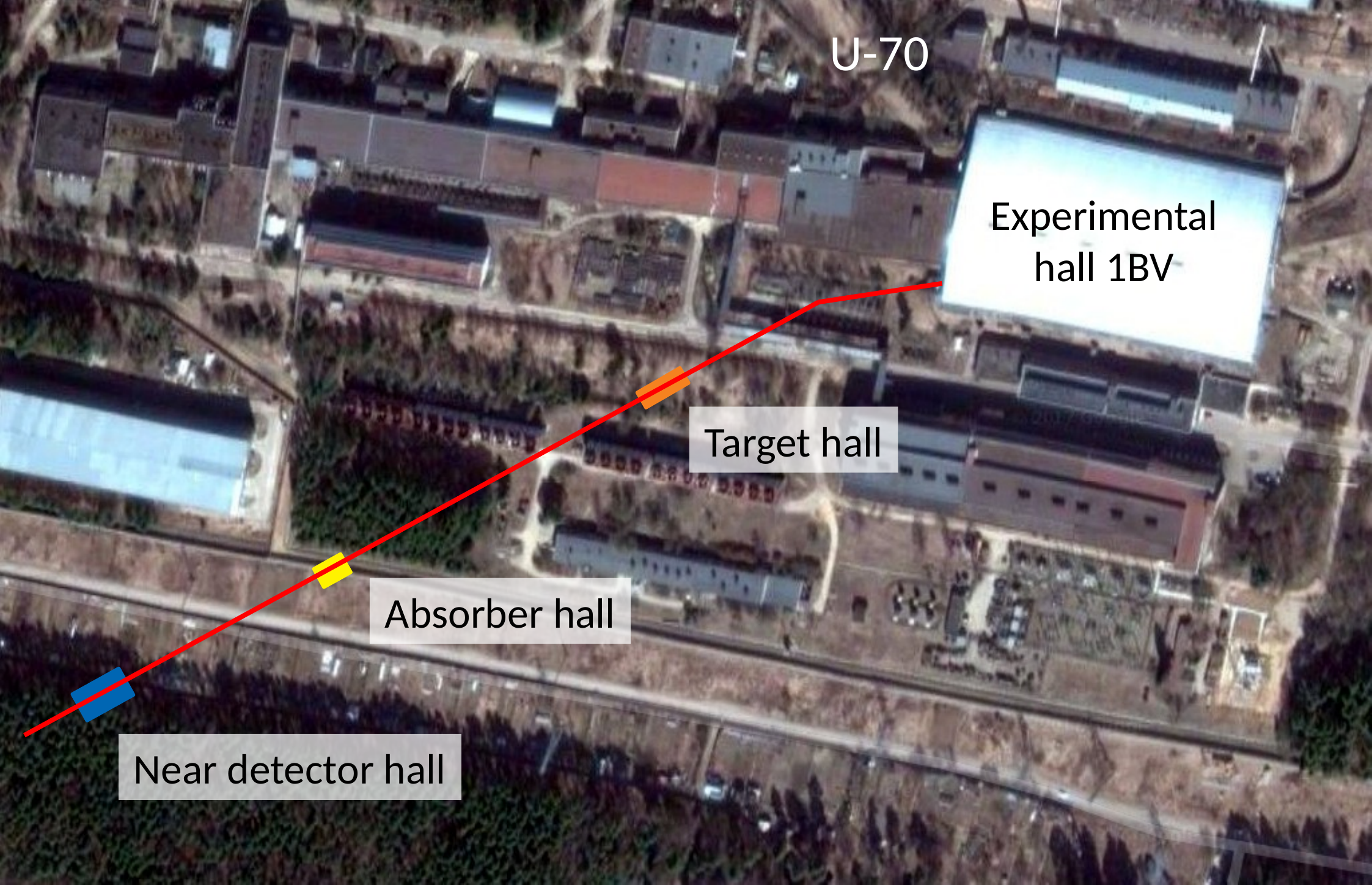}
    \includegraphics[width=14.1cm]{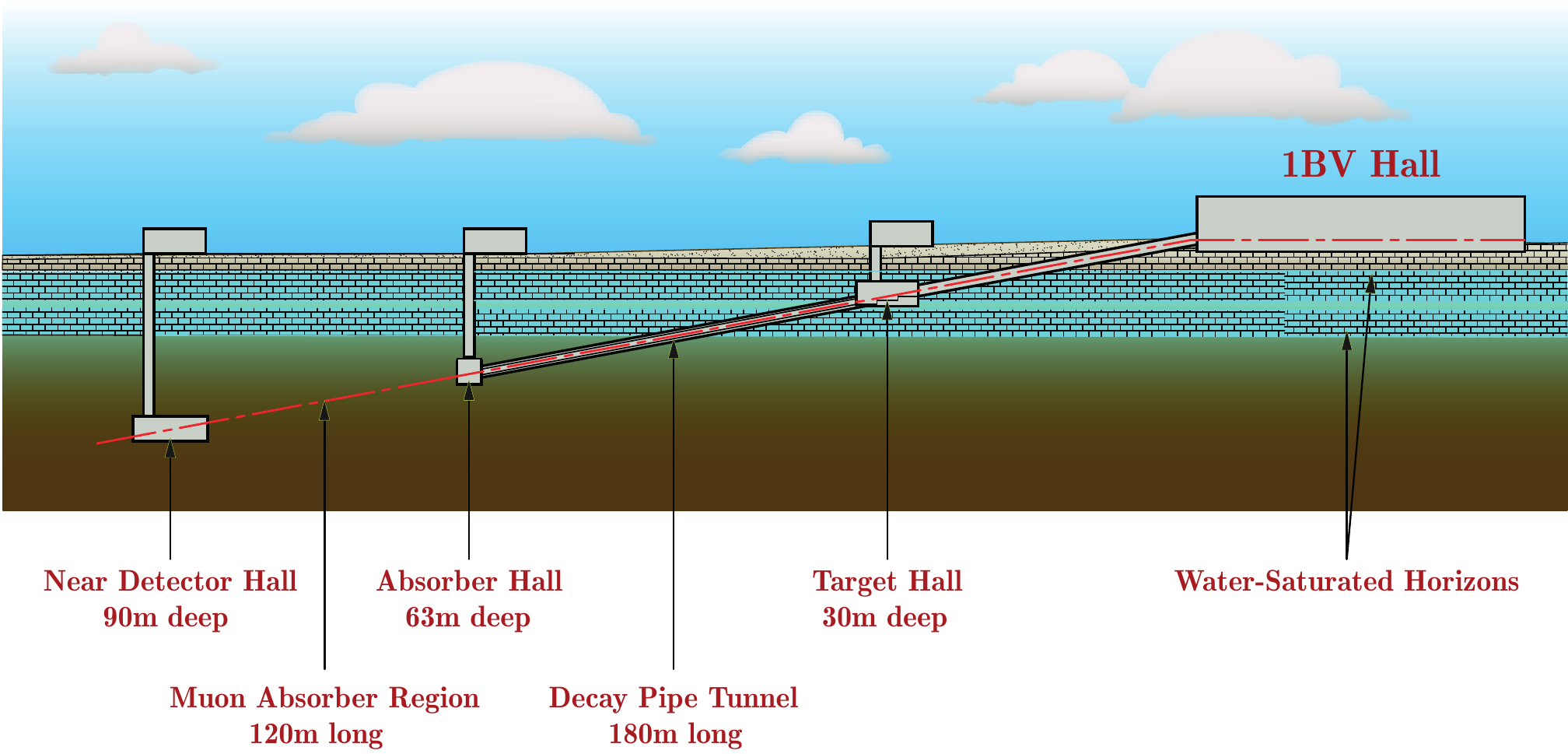}
  \caption{Top view and elevation view of the proposed neutrino beamline (the baseline design).
}
  \label{fig:p2o_beamline}
\end{figure}

A new neutrino beamline will need to be constructed at Protvino to enable the proposed research program.
In order to serve the P2O long-baseline experiment,
the beamline should be aligned towards the ORCA site (see Fig.~\ref{fig:p2o_beam_path}),
at an inclination angle of 11.7$^\circ$ (204~mrad) below the horizon.
A baseline design of the neutrino beamline, shown in Fig.~\ref{fig:p2o_beamline}, includes the following main components:
a beam extraction station, which could be installed on an accelerator section located in the main experimental hall;
a beam transport section, which delivers the primary protons from the extraction point to the target hall; a graphite target;
a secondary beam focusing system using magnetic horns;
a decay pipe, where neutrinos are produced from pion and kaon decays; and a beam absorber.
The longest section of the beamline is the decay pipe.
In the baseline design, the target hall is located at a depth of $\approx$ 30~m under ground level,
the decay pipe is $\approx$~180~m long (subject to optimization),
the absorber hall is $\approx$~63~m below ground level,
and the near detector hall is $\approx$~90~m below ground.
The magnetic horns will allow for reversal of the electric current polarity in order to choose between the neutrino and antineutrino mode.
Compared to the old neutrino beamline previously operated at Protvino, the new beamline design presents the following new challenges:
1) need for a higher beam intensity; 2) beamline to be constructed in an inclined tunnel.
These challenges are to be addressed in a dedicated R\&D study.

\begin{figure}
  \centering
    \includegraphics[width=8cm,trim={2.3cm 6.7cm 4.0cm 7.2cm},clip]{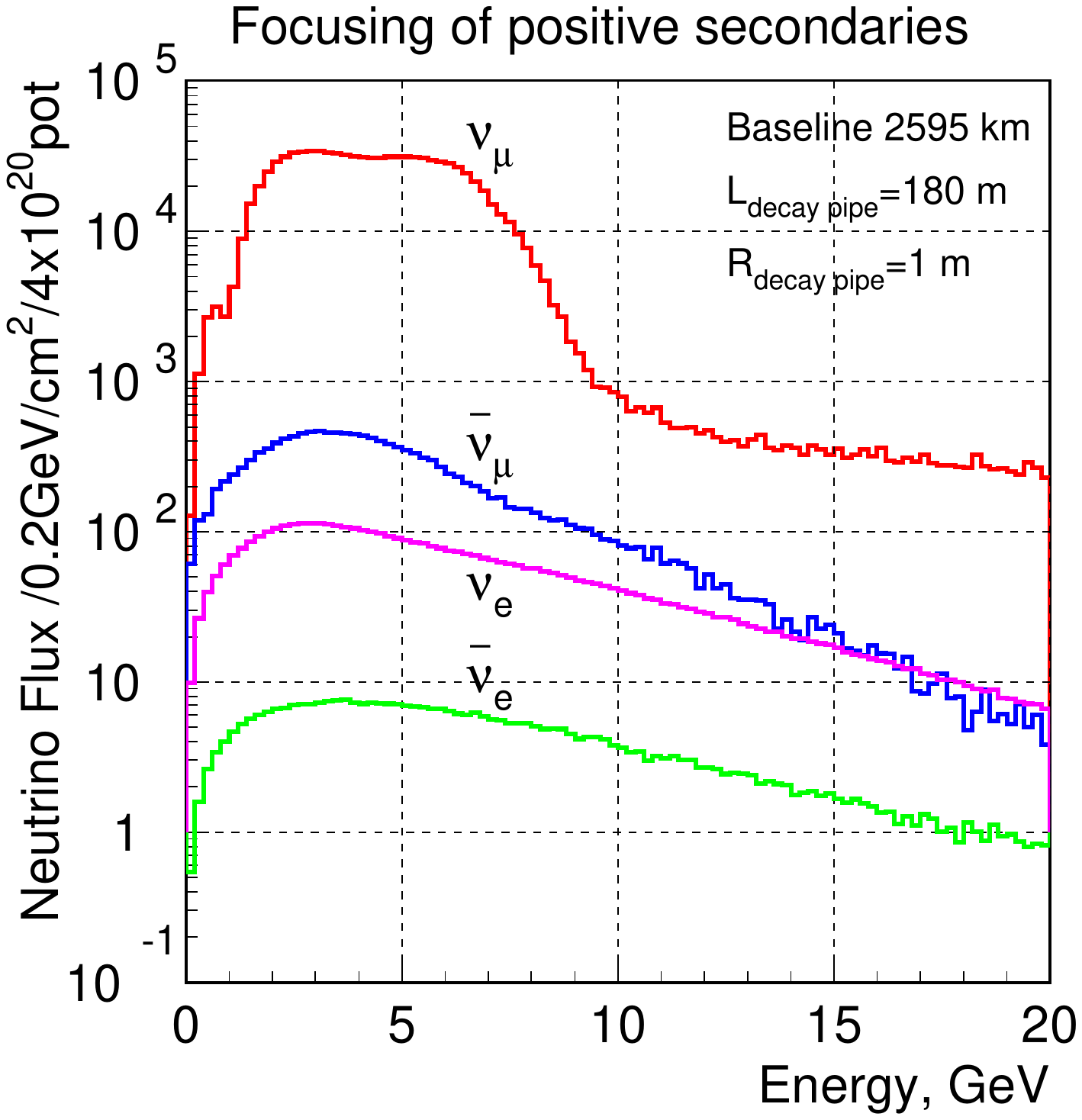}
    \includegraphics[width=8cm,trim={2.3cm 6.7cm 4.0cm 7.2cm},clip]{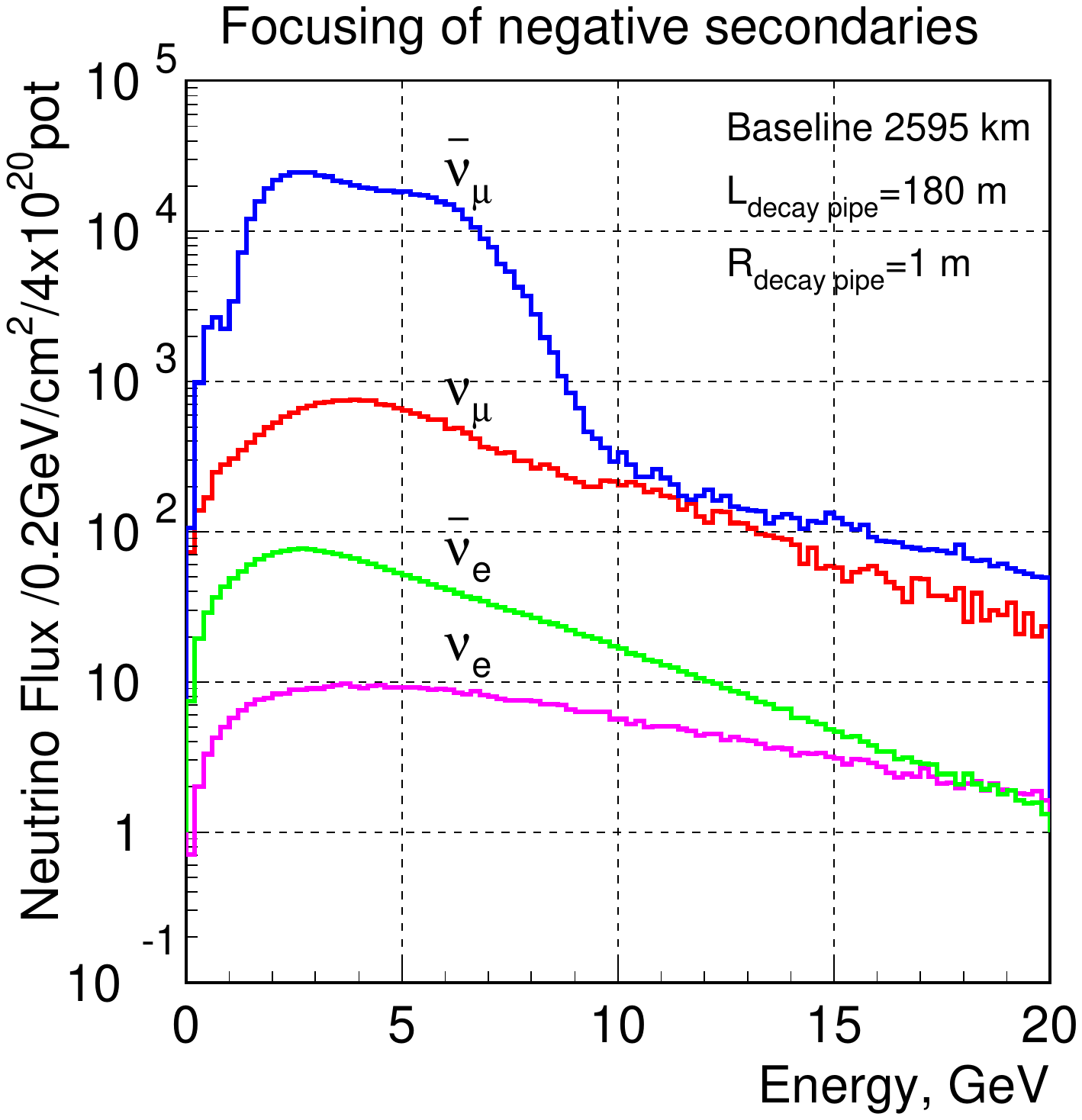}
  \caption{Simulated spectra of the P2O neutrino beam at the ORCA location in neutrino (left panel) and antineutrino (right panel) mode.
The absolute normalisation is given for $4 \times 10^{20}$ protons on target (POT),
which corresponds to 1 year of operation at the beam power of 450 kW, or 5 years with 90 kW.
}
  \label{fig:p20_beam_spectra}
\end{figure}

A relatively simple computer code was used to simulate the neutrino beam spectra of the proposed beamline, as described as follows.
Pions and kaons are generated in the target using analytical formulae for the fast calculation
of secondary particle yields in p-A interactions \cite{Bonesini2001}.
Decay weights and detector acceptances for neutrinos are calculated
at multiple locations as the particles are tracked along the beam line.
The neutrino spectra at the far detector site are computed
taking into account the angular distribution of the produced neutrinos
and assuming a zero off-axis angle. 
Absorption, scattering and energy loss of hadrons in the inner conductors
of the horns and in the decay pipe wall are taken into account,
but tertiary particles are not generated. 
This approach allows many variants to be checked at the preliminary stage of the
beamline design \cite{Abramov2002}.
This simplified approach may lead to a sizeable underestimation
of the fraction of $\bar\nu_e$ and $\bar\nu_\mu$ in the $\nu$ beam
($\nu_e$ and $\nu_\mu$ in the $\bar\nu$ beam),
but has only a small effect on the $\nu_e$ component of that beam
($\bar\nu_e$ component of the $\bar\nu_\mu$ beam).
Hence, for the $\nu_e$ ($\bar\nu_e$) appearance measurements considered in this paper,
this simplification appears adequate.
The obtained neutrino and antineutrino non-oscillated fluxes at the ORCA location are shown in Fig.~\ref{fig:p20_beam_spectra}.
As can be seen, the simulated set-up provides a high purity muon (anti)neutrino beam 
with a plateau in the neutrino energy distribution between 2 and 7~GeV.
A more detailed, full simulation study is planned for a future work.

\section{Near Detector}
\label{sect:near_detector}
Following the classic paradigm of long-baseline neutrino experiments, the primary purpose of the near detector is to monitor the energy spectrum, composition and direction of the neutrino beam close to the source, before the composition is modified by oscillations.
This is important for controlling the measurement uncertainties and thus achieving the targeted performance and sensitivity of the experiment.
The near detector can also be used for studies of neutrino-nucleus interactions, searches for short-baseline oscillations, and other studies.
The P2O near detector would be located $\sim$ 120~m downstream from the beam dump ($\sim$ 320~m from the proton target).
The detector should be large enough to fully contain hadronic cascades created by 5--10 GeV neutrinos.
Muon tracks exiting the main detector volume could be measured by additional muon detectors.
For reference, a 5 GeV muon travels $\approx$~22~m in water before stopping.

The choice of technology and materials for the near detector is a complex subject.
It is generally preferable to use the same material and detector technology for the near and far detector in order to reduce systematic uncertainties related to extrapolations from one target material to another, and from one detector technology to another.
However, additional considerations and constraints may call for other design choices.
For instance, the use of a higher granularity detector at the near site may be preferable, as it would allow for a more refined measurement of the neutrino interaction products, thus enabling more detailed studies of neutrino cross sections and related nuclear physics.
Constraints on the maximal dimensions of the near detector hall may call for use of heavy materials to reduce the detector dimensions.
The final design of the near detector needs to balance all requirements and constraints.
Several design options for the P2O near detector are currently under consideration.
They can be subdivided into two main groups:

1) A high granularity detector containing water in one or several of its subsystems.
This design option is inspired in part by the T2K's ND280 \cite{ND280_TDR} and NO$\nu$A near detector \cite{NOVA_TDR} designs.

2) A large water tank instrumented with PMTs.
This is similar to the TITUS and NuPRISM designs proposed for T2HK \cite{Scott2016}.
This design could incorporate KM3NeT PMTs as light sensors, thus closely mimicking conditions of the far detector (ORCA).

The use of a water-based liquid scintillator is under consideration as a possible alternative to pure water for both design options.
A part of the detector could be filled with heavy water, which would be useful for studies of nuclear effects and determination of cross sections on free protons and neutrons. 
The option to use several detectors with different measurement techniques can be considered as well.

\section{Science with the Neutrino Beam from Protvino to ORCA}
\label{sect:science_far_detector}

\begin{figure}
  \centering
    \includegraphics[width=10cm]{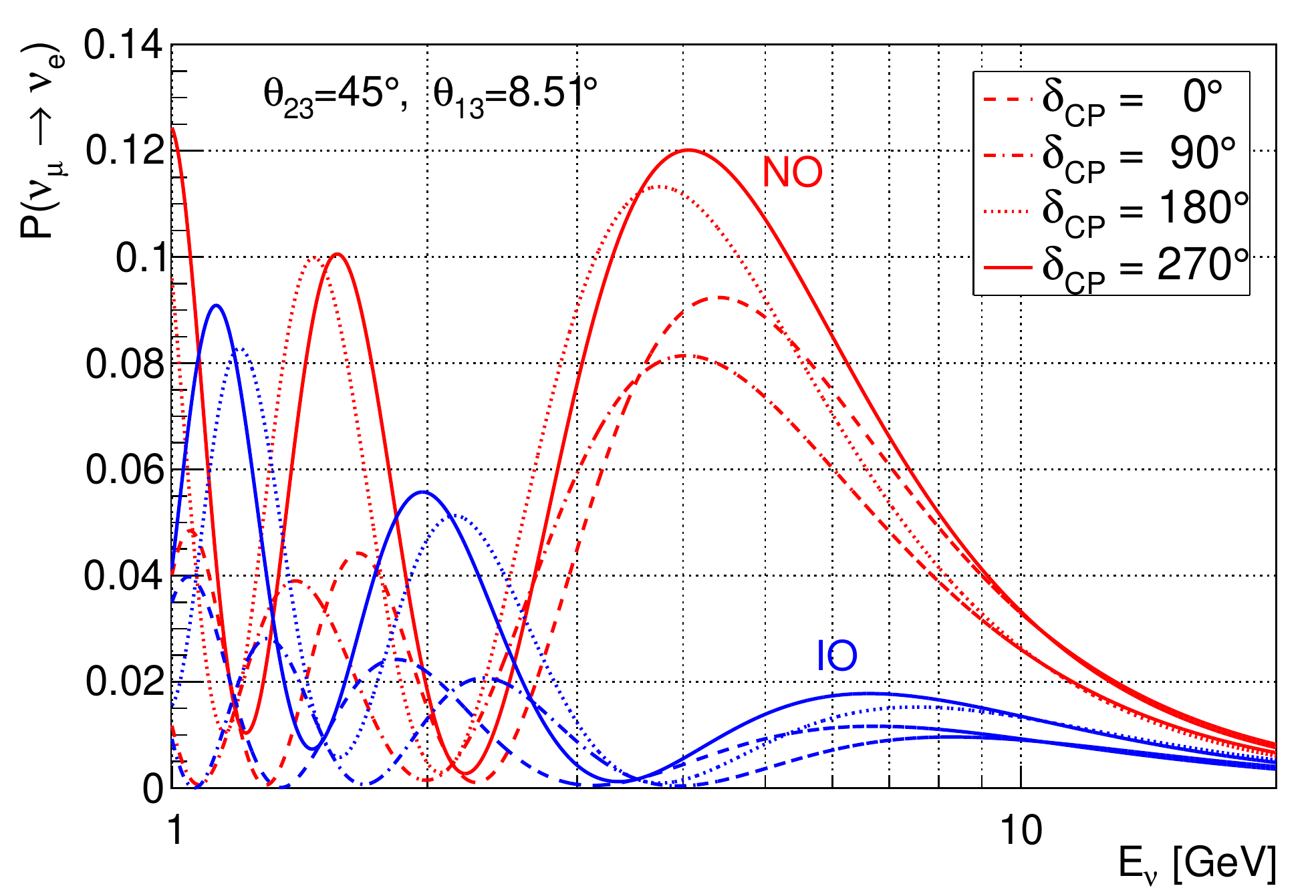}
  \caption{Oscillation probabilities for $\nu_\mu \rightarrow \nu_e$ (electron neutrino appearance) for a baseline of 2595 km for normal (NO) and inverted (IO) mass ordering.}
  \label{fig:oscprob}
\end{figure}

\begin{figure}
  \centering
    \includegraphics[width=11cm]{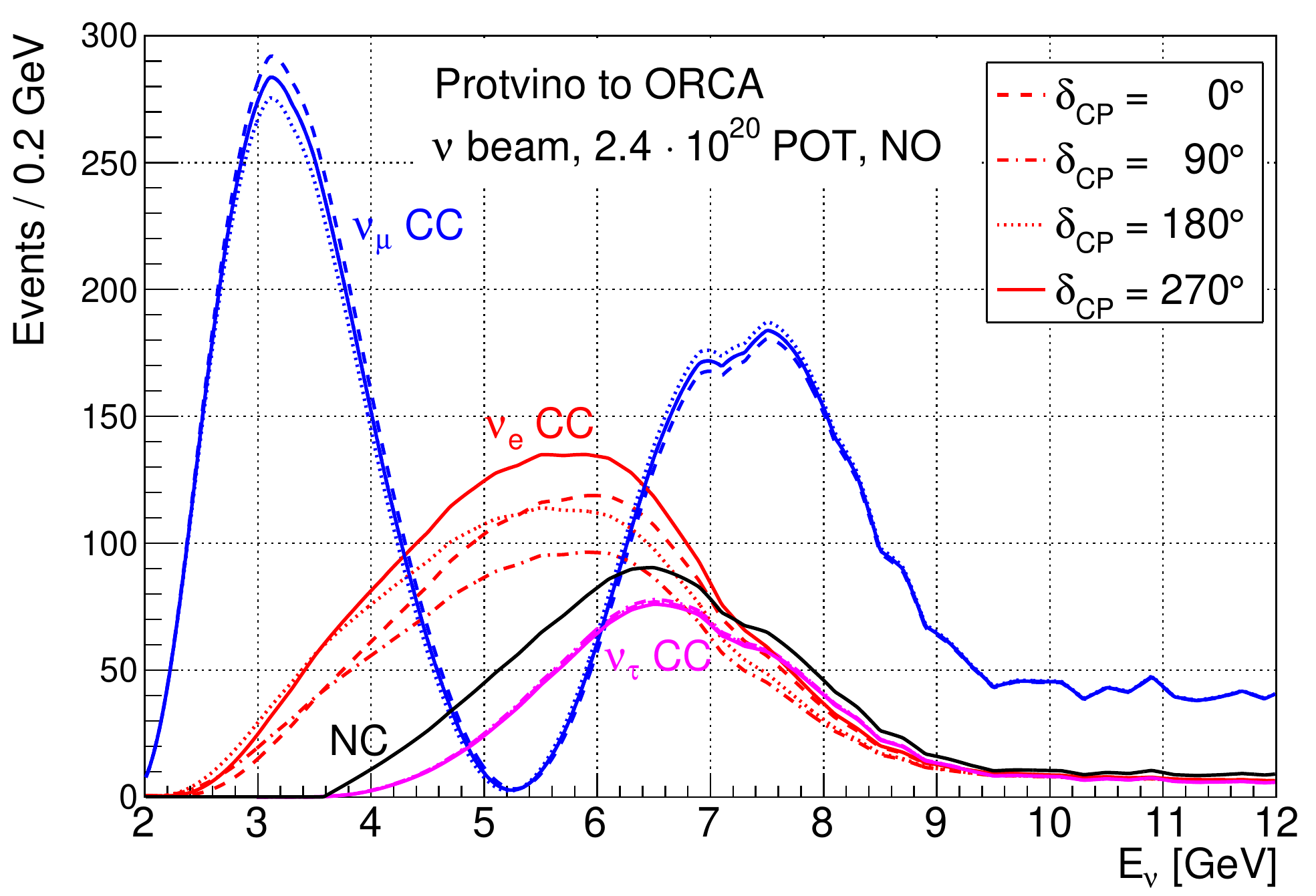}
    \includegraphics[width=11cm]{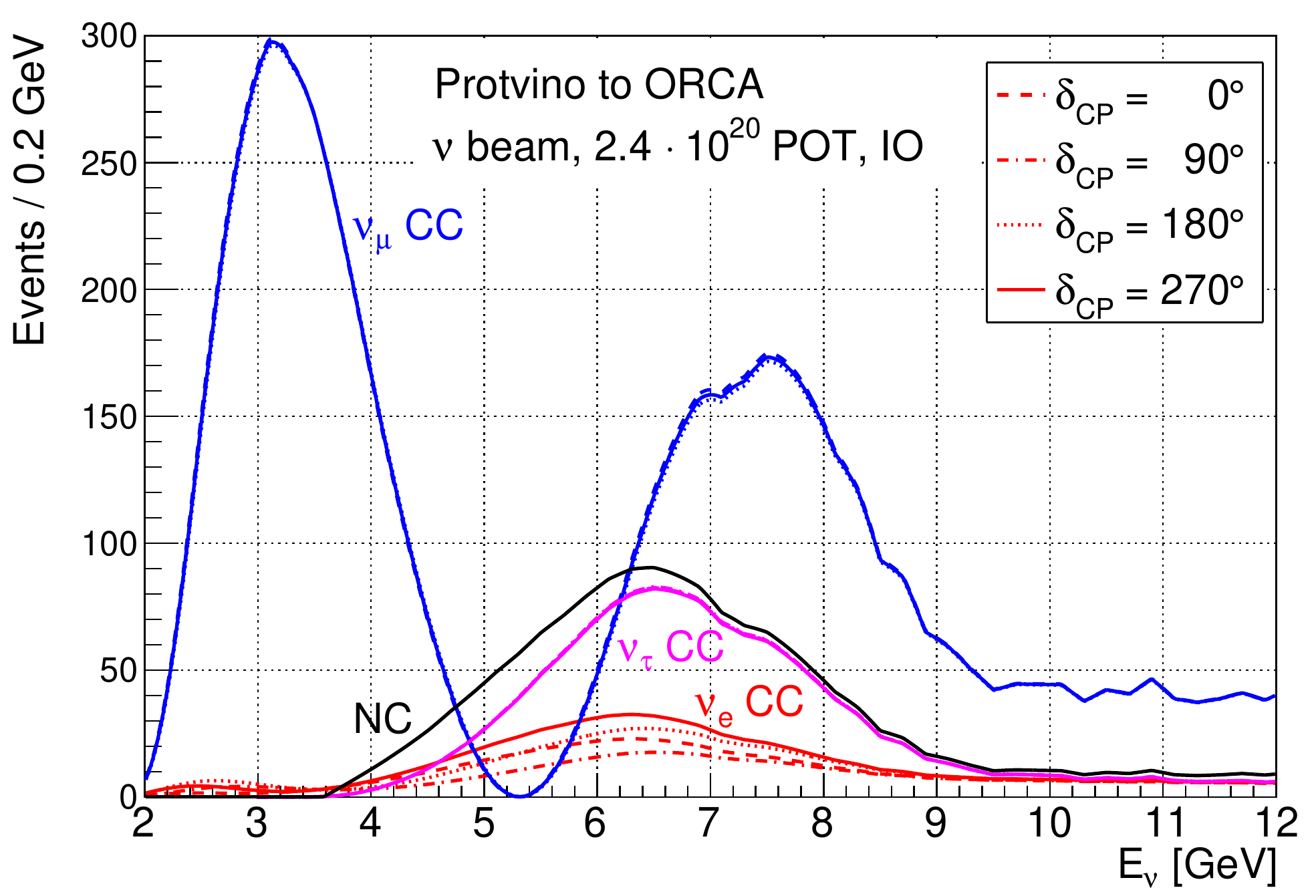}
  \caption{Energy distribution of the expected number of neutrino events that would be detected by ORCA after 3 years of running with a 90 kW beam from Protvino (neutrino mode) for 4 different values of the CP phase for the case of normal (top plot) and inverted (bottom plot) neutrino mass ordering.
$\theta_{23} = 45^\circ$ is assumed.
The x-axis shows the true neutrino energy.
}
  \label{fig:event_numbers}
\end{figure}

Sending a neutrino beam from Protvino to ORCA provides a baseline of 2595 km,
larger than any accelerator neutrino experiment currently operating or planned elsewhere.
The first $\nu_\mu \rightarrow \nu_e$ oscillation maximum is then at $E_\nu$ $\approx$ 5 GeV, within the energy range readily available from the U-70 synchrotron and within ORCA's nominal energy range.
In this energy regime, the neutrino interaction cross section is dominated by deep inelastic scattering, which is relatively well described theoretically (compared to resonant interactions which dominate at $\approx$ 2--3 GeV), thus facilitating high-precision measurements of neutrino flavour oscillations.
For reference, a recent study by the MINER$\nu$A Collaboration reported a 10\% uncertainty for the total neutrino cross section at 2.5 GeV and a 5\% uncertainty at 5 GeV \cite{MINERvA2017}.
The 2595 km baseline is well suited for probing the CP-violating Dirac phase $\delta_{\textrm{CP}}$, as well as for measuring the matter resonance effect ($E_{\textrm{res}}$ = 4 GeV for the Earth crust) \cite{Brunner2013,Vissani2013}.
The effects of the mass ordering and $\delta_{\textrm{CP}}$ are most pronounced in the $\nu_e$ appearance channel (see Figs.~\ref{fig:oscprob} and \ref{fig:event_numbers}).
The large instrumented volume of ORCA, 8 million cubic meters, will allow for the detection of thousands of neutrino events per year, even with a relatively modest accelerator beam power and despite the very long baseline.

\subsection{Sensitivity calculation procedure} 
To evaluate the scientific potential of the P2O experiment, its sensitivity to determine the neutrino mass ordering and to measure the effect of CP violation is studied.
The sensitivity calculation procedure uses the Asimov set method and is identical to the procedure described in Section 3.6.2 of \cite{KM3NeT_LoI}.
The detector response of the KM3NeT/ORCA detector, based on a detailed simulation and reconstruction framework for track and shower topologies, is directly taken from~\cite{KM3NeT_LoI}. The event selection has been optimised to suppress background from mis-reconstructed atmospheric muons as well as optical noise from radioactivity and bioluminescence. This data analysis pipeline had been developed for atmospheric neutrino studies and does not yet include any potential improvements due to the known arrival direction and timing of the neutrino beam.
The known arrival direction of the beam would be used to constrain the missing transverse energy of the neutrino events, potentially allowing to identify NC events.
The beam neutrinos arrive during short (5~$\mu$s) beam spills which should allow for the suppression of background by a factor of $\sim$ 10$^6$.

The detector response in terms of energy dependent effective mass (Figs.~69, 88, 90 of~\cite{KM3NeT_LoI}), energy resolution (Figs.~68, 91
of~\cite{KM3NeT_LoI}) and particle identification (Fig.~99 of~\cite{KM3NeT_LoI}) is parametrised and fed into the oscillation sensitivity
framework.
For each neutrino interaction channel (($\nu_e,\nu_\mu, \nu_\tau$) CC, $\nu$ NC, ($\bar\nu_e, \bar\nu_\mu, \bar\nu_\tau$) CC, $\bar\nu$ NC) and both detection topologies (track, shower) a full set of parametrised detector response functions is provided. Further, the neutrino beam spectra shown in Fig.~\ref{fig:p20_beam_spectra} and neutrino cross sections from GENIE~\cite{GENIE,GENIE_manual} are used.
Oscillation probabilities are computed with OscProb \cite{OscProb} and/or GLoBES \cite{GLoBES} (both codes leading to very similar results).
All results presented in this section assume running with the positive beam polarity only.

Systematic uncertainties on neutrino oscillation parameters, normalisations and energy scales are considered.
The complete list of parameters together with their used true values and priors is given in Table~\ref{tab:syst}.
\begin{table}
\begin{center}
\begin{tabular}{ | c | c | c | c | c | c | c | c | }
  \hline
Parameter  & prior \\ \hline
$N_\mu$    &     $1\pm0.05$ \\
$N_e$      &     $N_\mu$ \\
$N_\tau$   &     $1\pm0.10$ \\
$N_{\textrm{NC}}$   &     $1\pm0.05$ \\
$\theta_{13}$  &   $(8.51\pm0.15)^\circ$ \\
$\theta_{23}$  &   $(45.0\pm2.0)^\circ$ \\
$\Delta m_{32}^2$ $[10^{-3}$ eV$^2]$ &  $2.5 \pm 0.05$ \\
ParticleID skew    &     $1\pm0.10$ \\
$E_{\textrm{scale}}$ overall        & $1\pm0.03$ \\
$E_{\textrm{scale}}$ e/$\mu$ skew   & $1\pm0.03$ \\
$E_{\textrm{scale}}$ had/e skew & $1\pm0.03$ \\
  \hline
\end{tabular}
\caption{Systematic uncertainties and priors (see text).}
\label{tab:syst}
\end{center}
\end{table}
Here $N_x$ denotes the uncertainties of the CC event rates of flavour $x$ while $N_{\textrm{NC}}$ is the corresponding NC event rate. The NC and $\nu_\tau$ CC cross sections are assumed to be determined with the required precision at the planned near detector. Details of these important measurements will be worked out in follow-up documents. The coupling of $N_\mu$ and $N_e$ is justified by lepton universality (the muon mass can be neglected at neutrino energies relevant for P2O) and the percent-level beam contamination  with $\nu_e$.
The neutrino oscillation parameters $\theta_{23}$ and $\Delta m_{23}^2$ are constrained by the given Gaussian priors for studies of CP-related parameters while they are left unconstrained for the mass ordering determination. The ParticleID skew describes the uncertainty of the track/shower identification procedure while the different energy scale parameters $E_{\textrm{scale}}$ refer to systematic uncertainties in the energy measurements. The two energy scale skew parameters are used to allow for separate energy measurement scales for $\nu_e$, $\nu_\mu$ and hadronic channels (NC and $\nu_\tau$).
The choice of priors for the oscillation parameters $\Delta m_{32}^2$ and $\theta_{23}$ is motivated in part by recent results from global fits (see, e.g., \cite{Esteban2017}).
The $\theta_{13}$ prior refers to the recent measurement by Daya Bay \cite{DayaBay2018}.
The choice of values for the other priors is motivated by previous works,
including studies of ORCA sensitivity with atmospheric neutrinos \cite{KM3NeT_LoI}
as well as other long baseline experiments, in particular DUNE \cite{DUNE_CDR,DUNE_IDR}.
These choices will be refined in follow-up studies.

\subsection{Sensitivity to mass ordering and CP phase}
\begin{figure}
  \centering
    \includegraphics[width=11.0cm]{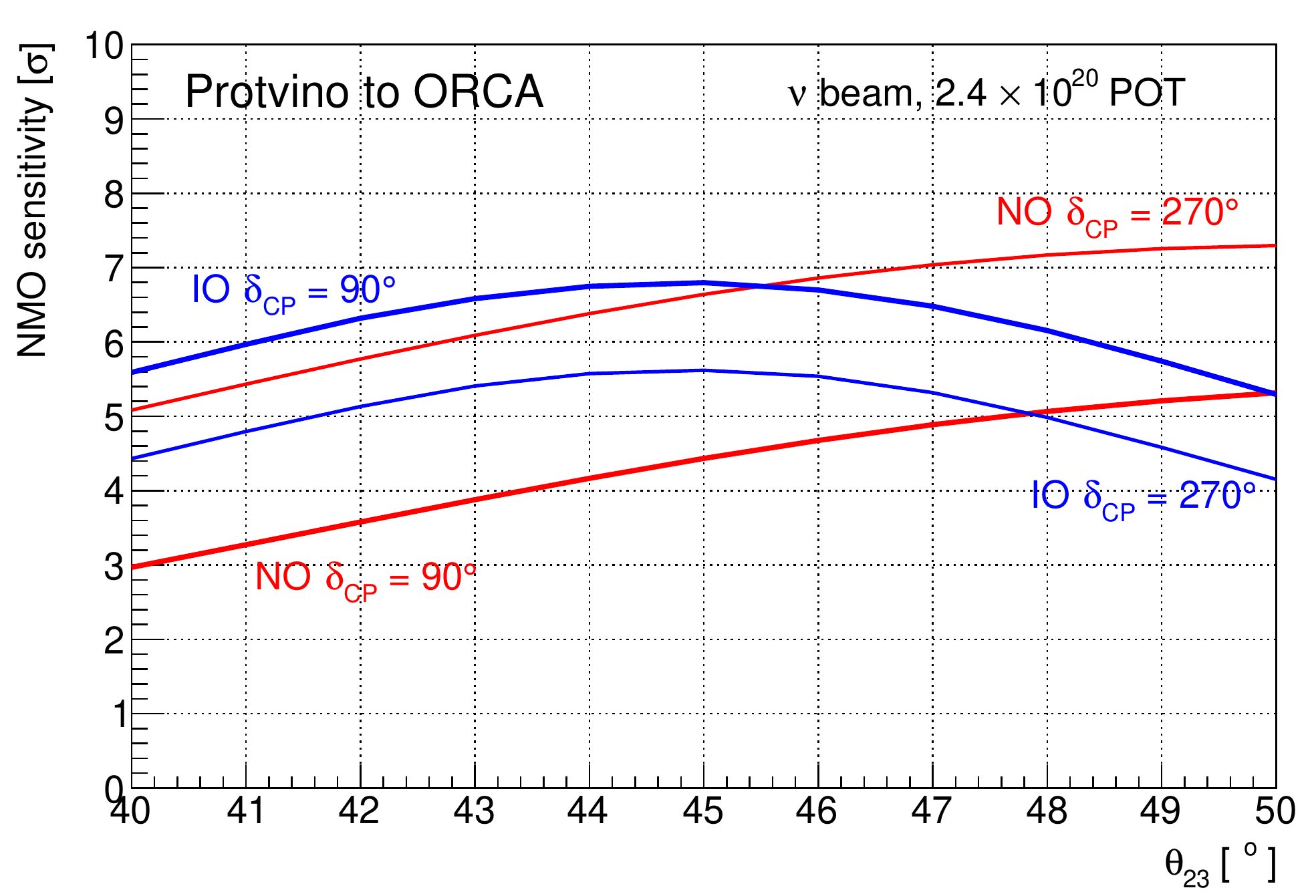}
  \caption{Sensitivity of P2O to the neutrino mass ordering (NMO) as a function of the $\theta_{23}$ mixing angle after 3 years of running with a 90 kW beam (positive beam polarity).
The $\theta_{23}$ and $\delta_{\textrm{CP}}$ values chosen provide the most and the least favourable scenarios for both normal (NO) and inverted mass ordering (IO).
One year of running with the 90 kW beam corresponds to $\approx 0.8 \times 10^{20}$ protons on target (POT).
}
  \label{fig:sensitivity_to_mass_hierarchy_1}
\end{figure}

\begin{figure}
  \centering
    \includegraphics[width=11.0cm]{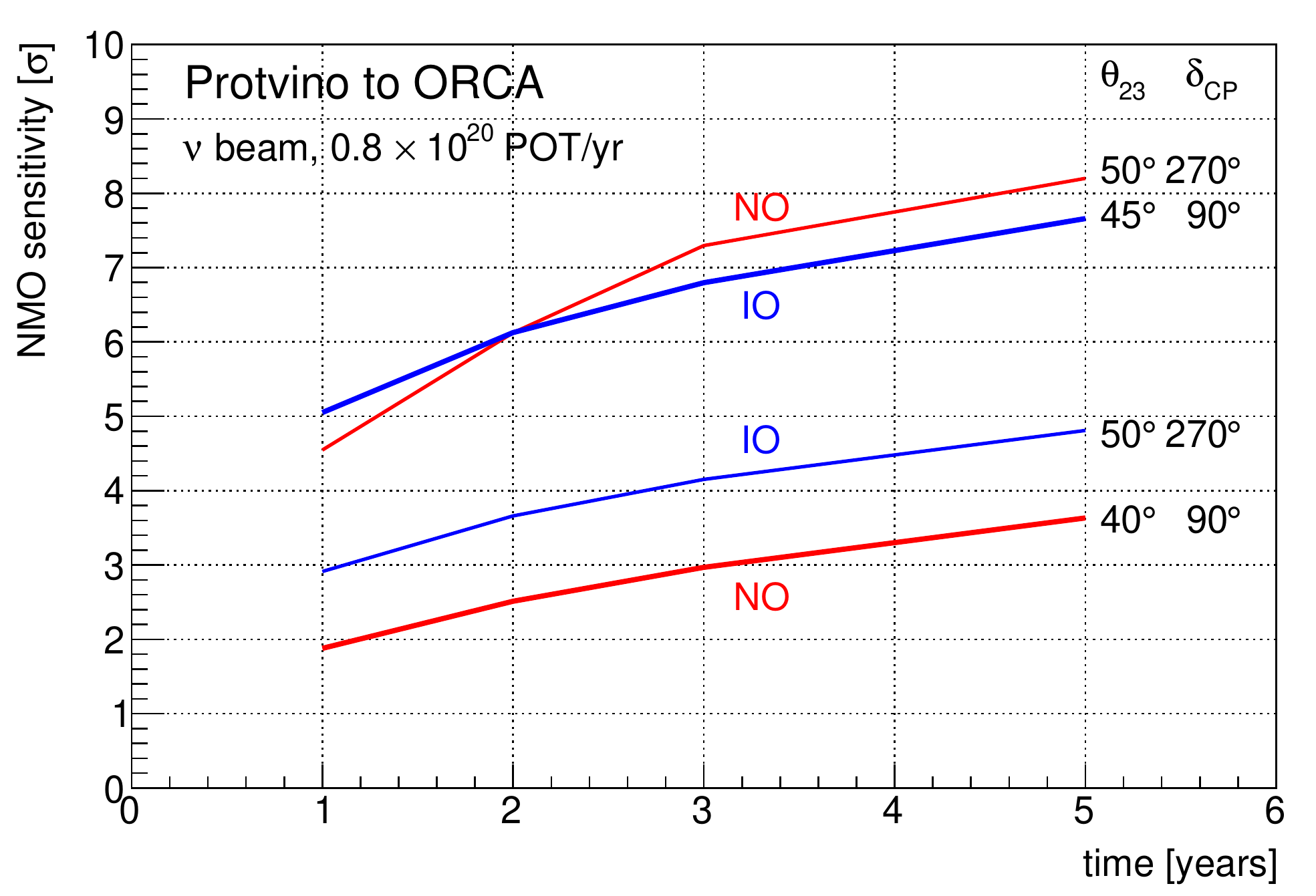}
  \caption{Sensitivity of P2O to neutrino mass ordering as a function of the accumulated exposure time with the 90 kW beam (positive beam polarity).
For both normal and inverted ordering, the most and the least favourable scenarios are shown.
}
  \label{fig:sensitivity_to_mass_hierarchy_2}
\end{figure}

With the procedure described above, the following results are obtained.
The neutrino mass ordering would be determined with a 4--8$\sigma$ statistical significance after one year of running with a 450~kW beam or after five years with a 90~kW beam (using positive beam polarity).
Three years of running with a 90~kW beam would already be sufficient to reach a $\ge$3$\sigma$ sensitivity, 
for any value of $\theta_{23}$ between 40$^\circ$ and 50$^\circ$ and any value of $\delta_{\textrm{CP}}$
(see Figs.~\ref{fig:sensitivity_to_mass_hierarchy_1},\ref{fig:sensitivity_to_mass_hierarchy_2}).
This would provide a solid confirmation of the $\approx$ 3--5$\sigma$ result expected to be achieved in the coming years by 
ORCA using atmospheric neutrinos, NO$\nu$A using accelerator neutrinos, and JUNO \cite{JUNO} using reactor neutrinos.
 
\begin{figure}
  \centering
    \includegraphics[width=11cm]{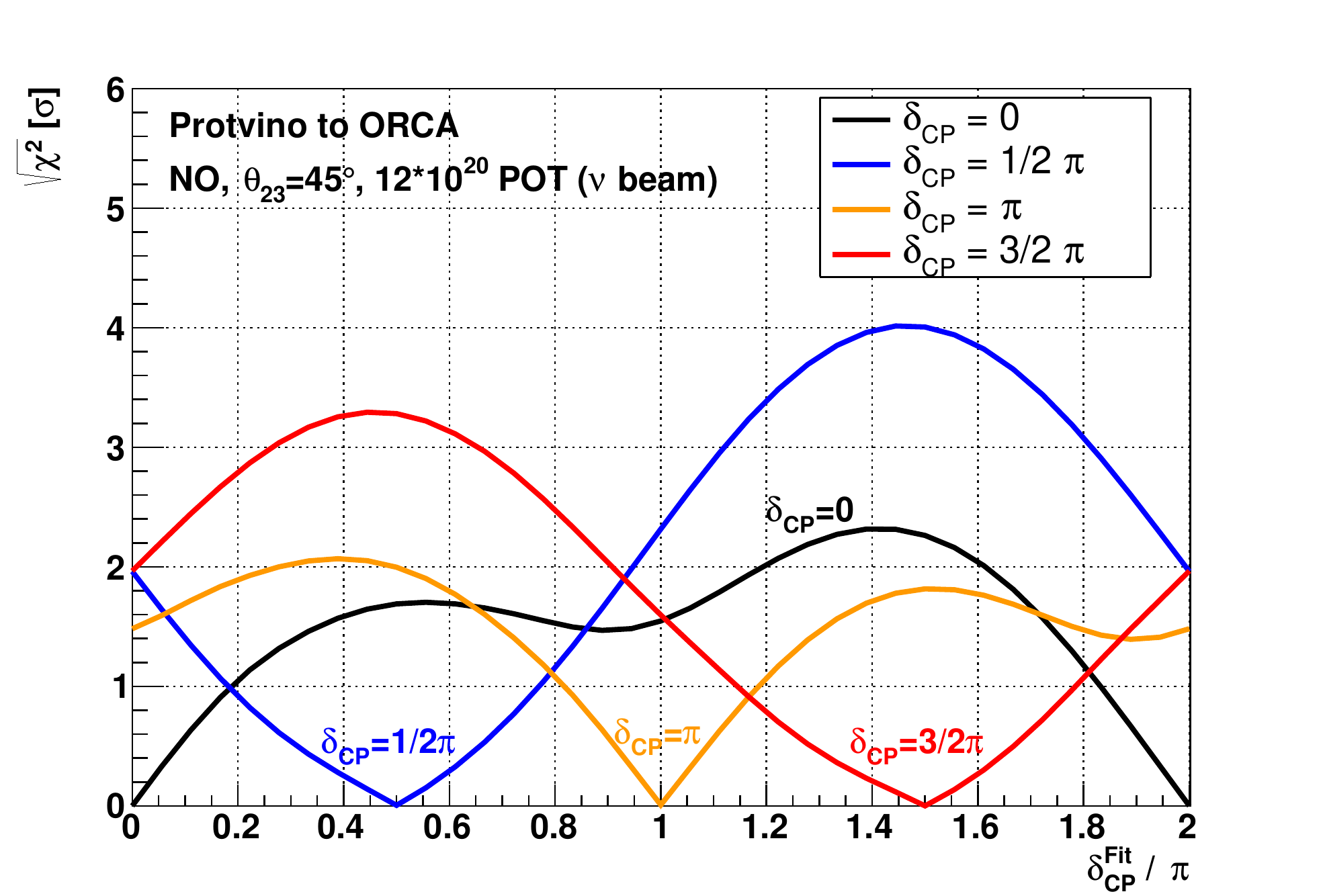}
  \caption{Sensitivity to exclude certain values of the CP phase $\delta_{\textrm{CP}}$ in the P2O experiment
after 3~years of running with a 450~kW beam (positive beam polarity) 
for 4 example values of the true $\delta_{\textrm{CP}}$ (0$^\circ$, 90$^\circ$, 180$^\circ$ and 270$^\circ$).
Each curve represents an average $\sqrt{\Delta\chi^2}$ plot that would be obtained if the true $\delta_{\textrm{CP}}$ value equals the example value indicated by the corresponding label.
$\sqrt{\Delta\chi^2}=1$ corresponds to the 1$\sigma$ uncertainty.}
  \label{fig:simulated_CP_measurements}
\end{figure}

After 3 years of operation with the 450 kW beam, the P2O experiment could achieve up to a 2$\sigma$ sensitivity to discover CP violation.
At the P2O baseline of 2595~km, most of the sensitivity to $\delta_{\textrm{CP}}$ comes from one beam polarity: positive for the case of normal mass ordering and negative for the case of inverted mass ordering.
Alternating between positive and negative beam polarities ($\nu$ and $\bar{\nu}$ modes) can help resolve the $\delta_{\textrm{CP}}$--$\theta_{23}$ degeneracy but otherwise does not necessarily improve the experiment sensitivity.
For that reason, the P2O sensitivity to $\delta_{\textrm{CP}}$ was derived assuming a fixed beam polarity chosen according to the mass ordering.
For the case of normal mass ordering, after 3 yr with the 450~kW beam (positive polarity),
the 1$\sigma$ accuracy on the value of $\delta_{\textrm{CP}}$ is of 30$^\circ$--60$^\circ$,
depending on the true $\delta_{\textrm{CP}}$ value (see Fig.~\ref{fig:simulated_CP_measurements}).
For the case of inverted mass ordering, a negative beam polarity will need to be used to obtain a measurement of $\delta_{\textrm{CP}}$.
In that case, reaching the same level of sensitivity to $\delta_{\textrm{CP}}$ will take a 2--3 times longer exposure time
(due to the lower production efficiency and interaction cross section of antineutrinos compared to neutrinos).

The systematic uncertainties have a relatively small effect on the mass ordering sensitivity.
For reference, doubling all of the detector-related uncertainties reduces the sensitivity shown in Figs.~\ref{fig:sensitivity_to_mass_hierarchy_1},\ref{fig:sensitivity_to_mass_hierarchy_2} at most by $\approx$ 0.7$\sigma$.
The systematic uncertainties play a more important role for the CP violation studies,
which rely on a high statistics measurement of a relatively small effect (as seen on Fig.~\ref{fig:event_numbers}).
The CP violation discovery potential of P2O becomes largely limited by the systematic measurement uncertainties already after a 3 year exposure to the 450 kW beam.
For reference, setting all the uncertainties to zero improves the $\delta_{\textrm{CP}}$ sensitivity shown in Fig.~\ref{fig:simulated_CP_measurements} threefold.

Another study, conducted independently and reported in \cite{Choubey2017}, finds similar results.
Minor differences with respect to our work can be explained by differences in the treatment of systematic uncertainties,
the choice of priors on the oscillation parameters, and the assumed beam spectra.

The estimated sensitivity of P2O to mass ordering and CP violation is compared to the sensitivity of some proposed and presently operating long-baseline experiments in Table~\ref{table1}.
Both T2K and NO$\nu$A have published experimental constraints on the mass ordering and CP violation \cite{T2K_2018,NOvA2018} which are within statistical error bars from the sensitivity figures given in Table~\ref{table1}.
The mass ordering sensitivity of P2O exceeds that of NO$\nu$A and is competitive with the sensitivity of DUNE.
T2K has a marginal sensitivity to mass ordering due to an insufficiently long baseline.
The CP violation sensitivity of P2O is competitive with T2K and NO$\nu$A.
Both T2K and NO$\nu$A alternate between using positive and negative beam polarity,
both polarities providing sensitivity to $\delta_{\textrm{CP}}$.
The sensitivity values given in the table for P2O are for 3 yr at 450 kW with positive beam polarity only.
The 90 kW positive-polarity beam will produce $\sim$ 4000 neutrino events in ORCA per year.
In the case of normal mass ordering, $\approx$ 700 of these events will be $\nu_e$ events.
For comparison, the DUNE experiment, using a 1.1~MW beam in combination with a 40 kt liquid argon detector over a 1300 km baseline, will detect $\approx$ 250 $\nu_e$ events per year.
\begin{table}
\footnotesize
\def\arraystretch{1.3}
\renewcommand\tabcolsep{3.5pt}
\begin{center}
\begin{tabular}{ | c | c | c | c | c | c | c | c | }
  \hline
  Experiment & \multicolumn{2}{|c|}{\bf T2K} &  {\bf T2HK} & {\bf NO$\nu$A}  & {\bf DUNE}  & \multicolumn{2}{|c|}{\bf P2O}  \\
  \hline
  Location & \multicolumn{2}{|c|}{Japan} & Japan & USA & USA & \multicolumn{2}{|c|}{Russia/Europe} \\
  \hline
  Status & \multicolumn{2}{|c|}{operating} & proposed & operating & construction & \multicolumn{2}{|c|}{proposed} \\
  \hline
  Accelerator facility & \multicolumn{2}{|c|}{J-PARC} & J-PARC & Fermilab & Fermilab & \multicolumn{2}{|c|}{Protvino} \\
  \hline
  Baseline & \multicolumn{2}{|c|}{295 km} & 295 km & 810 km & 1300 km & \multicolumn{2}{|c|}{2595 km} \\
  \hline
  Off-axis angle & \multicolumn{2}{|c|}{2.5$^\circ$} & 2.5$^\circ$ & 0.8$^\circ$ & 0$^\circ$ & \multicolumn{2}{|c|}{0$^\circ$} \\
  \hline
  1-st max $\nu_\mu \rightarrow \nu_e$ & \multicolumn{2}{|c|}{0.6 GeV} & 0.6 GeV & 1.6 GeV & 2.4 GeV & \multicolumn{2}{|c|}{4 GeV} \\
  \hline
  Detector & \multicolumn{2}{|c|}{SuperK} & HyperK & NO$\nu$A & DUNE & ORCA & Super-ORCA \\
  \hline
  Target material & \multicolumn{2}{|c|}{pure water} & pure water & LS & liquid Ar & \multicolumn{2}{|c|}{sea water} \\
  \hline
  Detector technology & \multicolumn{2}{|c|}{Cherenkov} &  Cherenkov & LS & TPC & \multicolumn{2}{|c|}{Cherenkov} \\
  \hline
  Fiducial mass & \multicolumn{2}{|c|}{22 kt} & 186 kt & 14 kt & 40 kt & 8000 kt & 4000 kt  \\
  \hline
  Beam power  & 500 kW &  & 1300 kW & 700 kW & 1070 kW & 450 kW & 450 kW \\
  \hline
  $\nu_e$ events per year (NO) & $\sim$ 20 & & 230 & $\sim$ 20 & 250 & 3500 & 3400 \\
  \hline
  $\bar{\nu}_e$ events per year (IO) & $\sim$ 6 & & 165 & $\sim$ 7 & 110 & 1200 & 1100 \\
  \hline
  NMO sensitivity ($\delta_{\textrm{CP}} = \pi/2$)  &  -           & -          & 4$\sigma$ & 1$\sigma$ & 7$\sigma$ & 8$\sigma$ & $>$ 8$\sigma$ \\
  \hline
  CPV sensitivity ($\delta_{\textrm{CP}} = \pi/2$) & 1.5$\sigma$ & 3$\sigma$ & 8$\sigma$ & 2$\sigma$ & 7$\sigma$ & 2$\sigma$  & 6$\sigma$ \\
  \hline
  1$\sigma$ error on $\delta_{\textrm{CP}}$ ($\delta_{\textrm{CP}} = \pi/2$) & & & 22$^\circ$ & & 16$^\circ$ & 53$^\circ$ & 16$^\circ$ \\
  \hline
  1$\sigma$ error on $\delta_{\textrm{CP}}$ ($\delta_{\textrm{CP}} = 0$) & & & 7$^\circ$ & & 8$^\circ$ & 32$^\circ$ & 10$^\circ$ \\
  \hline
  Year / data taking years & 2018 & 2026 & 10 yr & 2024 & 10 yr & 3 yr & 10 yr \\
  \hline
  Refs. & \cite{T2K_2018} & \cite{T2K_2016} & \cite{T2HK_2015,T2HK_2018} & \cite{NOVA_TDR,NOvA2018b} & \cite{DUNE_CDR,DUNE_IDR} &  &  \\
  \hline
\end{tabular}
\caption{Sensitivity of present and future long-baseline accelerator neutrino experiments to neutrino mass ordering (NMO) and leptonic CP violation (CPV).
All sensitivities are given for the case of normal mass ordering.
Expected number of $\nu_e$ ($\bar{\nu}_e$) events per year is given for the case of normal (inverted) mass ordering using positive (negative) polarity beam.
LS stands for liquid scintillator.
Ten years for DUNE corresponds to 500 kt$\cdot$\,MW\,$\cdot$\,yr.
}
\label{table1}
\end{center}
\end{table}
A combined analysis of the atmospheric and accelerator neutrino data collected by ORCA will be possible,
improving  the systematic uncertainties and parameter degeneracies.

\section{Science with the Near Detector}
\label{sect:science_near_detector}

\begin{figure}
  \includegraphics[width=\linewidth]{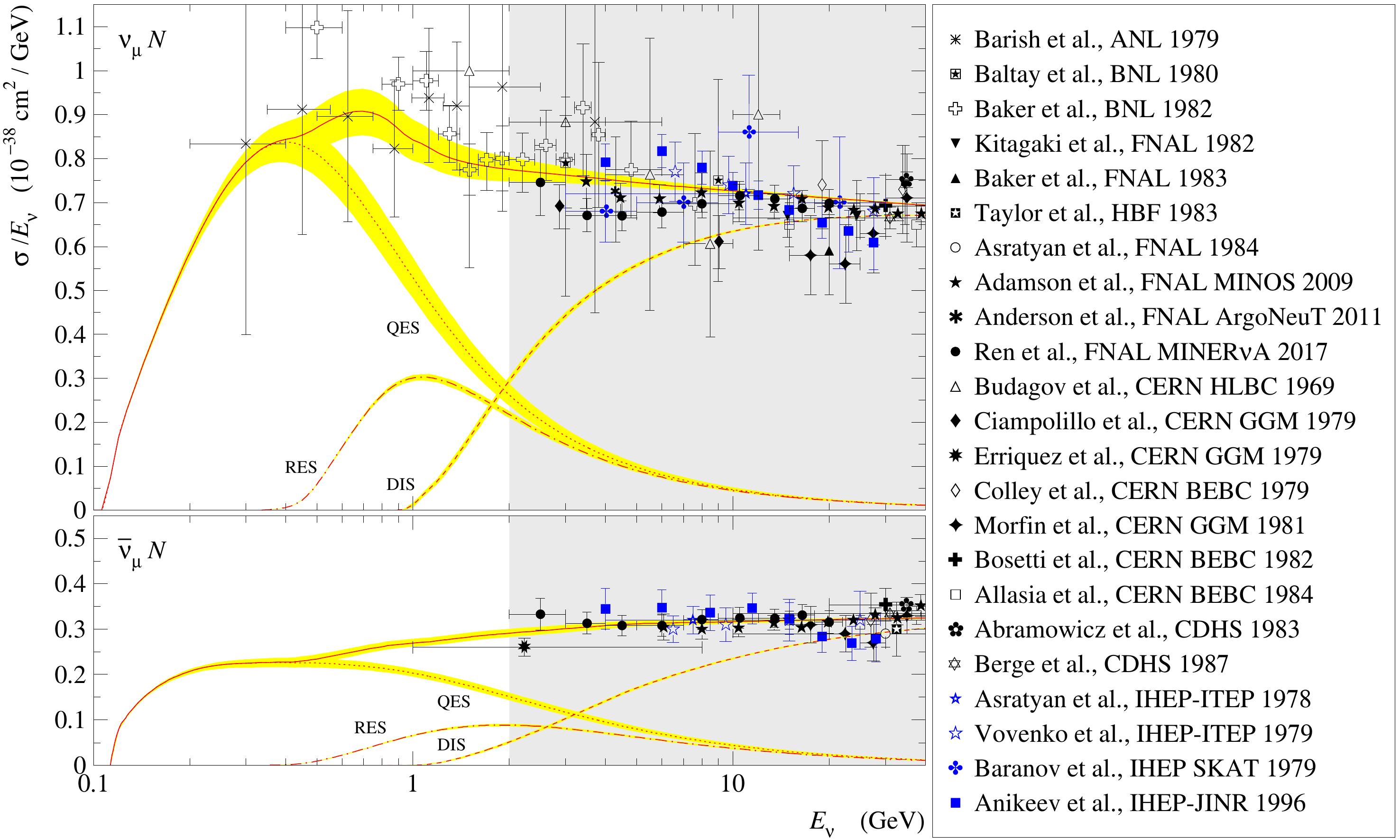}
  \caption{Total CC cross sections as functions of (anti)neutrino energy and
           normalized to energy for $\nu_\mu$ and $\overline{\nu}_\mu$
           scattering off isoscalar nucleons
           in comparison with experimental data~\cite{Experiments}.
           The curves and bands of theoretical uncertainties
           show the QES~\cite{QES} (with contribution from light strange hyperons
           productions in the case of $\overline{\nu}_\mu$ reactions~\cite{QHP}),
           RES~\cite{RES},
           and DIS~\cite{Kuzmin:2005} (see references therein)
           contributions and their sums.
           The shaded band indicates the energy range relevant for ORCA.}
  \label{Fig:sSUM}
  \end{figure}

  One of the main sources of systematic uncertainties
  in modern and future experiments for the study of fundamental properties of neutrinos
  is the uncertainty in the knowledge of the cross sections
  for neutrino and antineutrino interactions with nuclei.
  The cross sections due to charged and neutral currents
  are usually assumed to be a sum of cross sections for
  the reactions of (quasi)elastic (QES, ES) scattering,
  nucleon and baryon resonances production with their subsequent decay into a nucleon and pions (RES),
  production of kaons, light strange hyperons (for $\overline{\nu}_\mu$), and charmed mesons,
  and production of multiple hadrons including strange and charmed particles 
  in deep inelastic scattering (DIS);
  see Refs.~\cite{Kuzmin:2005,Tzanov:2010,Formaggio:2012,Gazizov:2016} and references therein.
  At neutrino energy range around 1 GeV,
  the cross sections for (Q)ES, RES, and DIS
  are comparable in magnitude (see Fig.~\ref{Fig:sSUM}).
  Current uncertainties in the theoretical calculation of the cross sections are
  related to difficulties in accounting for nontrivial nuclear effects 
  (meson exchange currents, exchange of baryon resonances between nucleons in the nucleus,
  multinucleon correlations, etc.)
  and significant uncertainties in the knowledge of the elastic and transition form factors of the nucleon,
  especially for the axial-vector and pseudoscalar, as well as for the nonstandard scalar and tensor form factors
  (for the latter two, at present, there are only very rough experimental upper limits).
  In the absence of a generally adopted and reliable model
  for neutrino-nucleus interactions
  which would be available in a wide energy range,
  different authors use different phenomenological models
  tuned to different energy ranges and detector targets.
  As a result, the values of the fundamental phenomenological parameters 
  for neutrino-nucleon interactions, extracted from the experiments,
  strongly depend on the interaction model used in analyses,
  and on average energies of neutrino and antineutrino beams
  (see, e.g., recent reviews \cite{Mahn:2018mai,Betancourt:2018bpu} and references therein).
  This in turn leads to uncertainties in extrapolations of the cross section models
  from one target material to another.

  High precision measurements with P2O will require an accurate knowledge of the
  (anti)neutrino cross sections in water.
  So far, the only experimental result on neutrino cross sections on a water target
  was obtained with the T2K experiment~\cite{Abe:2017rfw} at the mean neutrino energy $\sim$~1 GeV.
  Additional measurements appear necessary,
  both to improve the neutrino-nucleus interaction models and facilitate high-precision
  neutrino oscillation studies with P2O.
  The P2O near detector could provide a measurement
  of the neutrino and antineutrino cross sections
  with nucleons of a water target at neutrino energies from $\sim$~2 to 20~GeV.
  The obtained cross section data would also help to enhance the precision of the ORCA
  measurements using atmospheric neutrinos.

  The P2O near detector could be designed so as to allow for simultaneous measurements
  of the cross sections on two or more different nuclear targets,
  e.g. water and a carbonaceous scintillator.
  This would permit an unbiased comparison between the different materials, and, ultimately,
  a better understanding of the physics of neutrino scattering on nucleons bound in nuclei.
The cross section measurement programme could be further enhanced by additional specialized experiments.
In this context it is worth noting that a strong motivation exists
for a new experiment using the simplest targets, namely hydrogen and/or deuterium, 
in which case the investigation of the nucleon is separated from
the complications that arise due to in-medium nuclear effects.

%

The layout of the near detector has not yet been determined. But independently of its layout it will provide a large data sample of well measured neutrino interactions in the energy range from 3 to 8 GeV.
The flight path of 300~m from the proton target to the near detector corresponds to the first oscillation maximum for 2.4~GeV neutrinos at $\Delta m^2 = 10$~eV$^2$.
This would allow for an independent test of the high $\Delta m^2$ part of the so-called LSND anomaly \cite{LSND2001}
and a similar anomaly reported recently by the MiniBooNE Collaboration \cite{MiniBooNE2018}.
Both of these anomalies have been hypothesized to be caused by transitions to sterile neutrino states in the eV-scale mass range. 
Testing them in an U-70 neutrino beam has been suggested earlier~\cite{Garkusha2015}.

\section{Future Beyond ORCA}
\label{sect:future}

\catcode`\@=11 
\def\parenbar{\mathpalette\p@renb@r}
\def\p@renb@r#1#2{\vbox{%
		\ifx#1\scriptscriptstyle \dimen@.7em\dimen@ii.2em\else
		\ifx#1\scriptstyle \dimen@.8em\dimen@ii.25em\else
		\dimen@1em\dimen@ii.4em\fi\fi \offinterlineskip
		\ialign{\hfill##\hfill\cr
			\vbox{\hrule width\dimen@ii}\cr
			\noalign{\vskip-.3ex}%
			\hbox to\dimen@{$\mathchar300\hfil\mathchar301$}\cr
			\noalign{\vskip-.3ex}%
			$#1#2$\cr}}}
\catcode`\@=12 
\def\nuan{\parenbar{\nu}\kern-0.4ex}

A more densely-instrumented version of the ORCA detector, called Super-ORCA, is under discussion as a possible next step after ORCA. The Super-ORCA detector would provide a lower energy threshold for neutrino detection, better neutrino flavour identification capability and better energy resolution compared to ORCA. Such an upgrade would substantially enhance the scientific potential of the experiment, in particular the accuracy of the CP phase measurement.

For Super-ORCA, a 10 times denser detector geometry compared to ORCA is assumed along with a 4\,Mt fiducial volume.
This detector geometry has originally been studied for measuring the CP phase using $\sim$~GeV atmospheric neutrinos \cite{Hofestadt2018,Bruchner2018}, and has not been optimised for a neutrino beam from Protvino.

\subsection{Super-ORCA Detector Performance}
\label{sec:superoraca_detector_performance}

The expected detector performance of Super-ORCA has been estimated based on full event reconstruction applied to a simplified detector response simulation. The neutrino interaction (GENIE 2.10.2), particle propagation as well as Cherenkov photon generation and tracking in seawater is fully simulated using a similar simulation framework as described in \cite{ORCA_Intrinsic_Limits}. An up-to-date model of optical properties of the deep-sea water and optical background from ${}^{40}$K decays is taken into account.

Instead of a full detector simulation of a specific Super-ORCA detector geometry with multiple PMTs in DOMs along vertical strings, a generalised and simplified detector response is simulated.
In this simplified detector response, photons are randomly detected according to their wavelength-dependent detection probabilities ignoring the specifics of the detector geometry (such as partially contained events, and that PMTs are located in clusters, i.e. optical modules, with specific distances between them).
It has been verified based on a full detector simulation using KM3NeT/ORCA tools that only a small fraction of the DOMs further away than $\sim 20$\,m from the event detect multiple photons on the same DOM, so that it can be assumed that the clustering of PMTs in DOMs can be neglected for this study given the fine-grained instrumentation of Super-ORCA. Detected photons closer than 20\,m are not used for event reconstruction in the simplified detector response.
With these assumptions the detector response depends only on the instrumentation density. The assumed 10 times denser instrumentation than the ORCA detector,
corresponding to 115k 3-inch KM3NeT-PMT per Mt,
results in about $100$ detected photons per GeV for electromagnetic showers.

The simulated Cherenkov signatures are reconstructed with a full likelihood reconstruction assuming an electron ($e$) or muon ($\mu$) particle hypothesis plus an hadronic (had) shower hypothesis. The $\nu_e$/$\nu_\mu$ separation is mainly based on the likelihood difference of the fitted $e$+had and $\mu$+had event hypotheses. The different angular profiles of the emitted Cherenkov light can be exploited for $e$/$\mu$ separation. Due to the large photon scattering length in deep-sea water ($\lambda_s^{\rm eff} = \lambda_s / \left [ 1 - \langle \cos (\theta_s) \rangle \right ] \approx 265$\,m for a photon wavelength of $470$\,nm, see \cite{TransmissionOfLight}), the light emission characteristics are conserved over sufficiently large distances, so that information from a large detector volume (large lever arm) can contribute to event reconstruction, resulting in good
direction resolutions and $e$/$\mu$ separation capabilities.

\begin{figure}
  \centering
  \includegraphics[width=11.0cm]{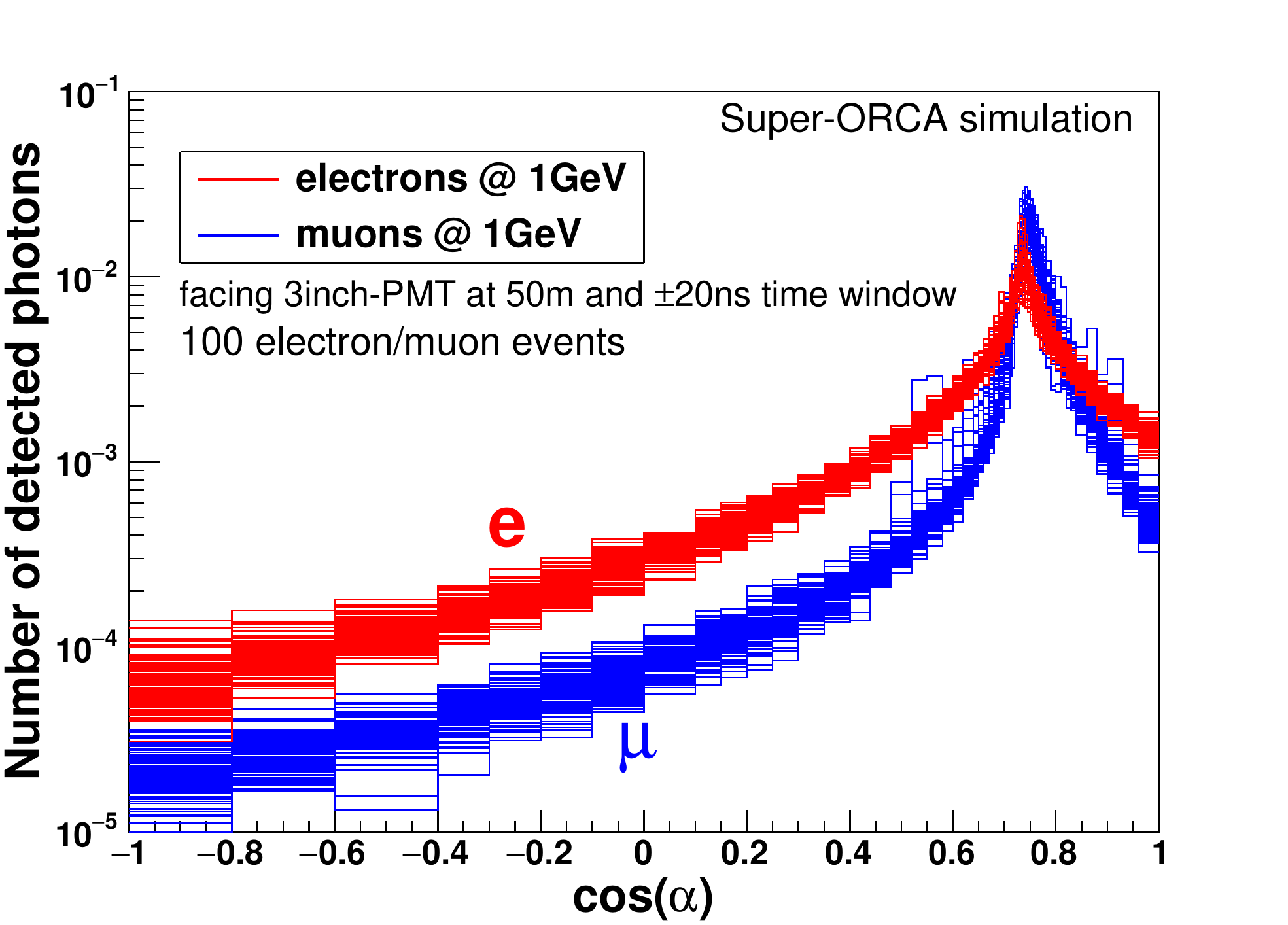}
  \caption{
  Angular profile of the Cherenkov light recorded 50\,m away from simulated $1$\,GeV electrons and muons. 
  Each line shows one out of 100 different simulated electron and muon events. 
  The number of detected photons per 3-inch PMT pointing towards the simulated particle as a function of the cosine of the angle $\alpha$ between the initial particle direction and the vector from the PMT to the particle position is shown.
  The particle position is defined as the barycentre of all Cherenkov photon emission positions.
  A $\pm$\,20\,ns integration time window around the average expected photon arrival time is used.
  Occasional outliers for muons are caused by large-angle scattering.
	   }
  \label{fig:superorca_agularProfile_emu1GeV}
\end{figure}

Figure~\ref{fig:superorca_agularProfile_emu1GeV} shows the angular profile of the Cherenkov light recorded 50\,m away from simulated $1$\,GeV electrons and muons.
Compared to muons, the angular profile is broader for electrons, leading to fuzzier Cherenkov cones. The same feature is used for $e$/$\mu$ separation in Super-Kamiokande \cite{SKflavourID}, where the fuzziness of the Cherenkov rings is exploited. 

The resulting Super-ORCA detector performance can be found in \cite{Hofestadt2018}. With the increased instrumentation density compared to ORCA, the energy threshold for neutrino detection is reduced to $\sim$~0.5~GeV, and the $\nu_\mu/\nu_e$ separation via the fuzziness of the Cherenkov cones allows for a selection of 95\%-pure samples of muon-like (dominated by $\nu_\mu$~CC) and electron-like events (dominated by $\nu_e$~CC). The neutrino energy resolution is $\approx$ 20\% at $E_\nu > 1$ GeV and is dominated by fluctuations in the number of emitted photons in the hadronic shower \cite{ORCA_Intrinsic_Limits}.

\subsection{Measuring CP phase $\delta_{\textrm{CP}}$ with Super-ORCA}
\label{sec:superoraca_sensitivity}
The $\delta_{\textrm{CP}}$ measurement with a neutrino beam from Protvino to Super-ORCA profits from running in neutrino as well as in antineutrino beam mode in order to resolve the $\delta_{\textrm{CP}}-\theta_{13}-\theta_{23}$ degeneracy  \cite{Barger2001,Choubey2017}. It is assumed that 50\% of the total exposure comes in neutrino beam mode and 50\% in antineutrino beam mode. An equal share between neutrino and antineutrino data was found to be close to optimal. The neutrino beam spectra shown in Figure~\ref{fig:p20_beam_spectra} are used and a beam power of 450\,kW is assumed.
For the sensitivity calculation, the same systematics and priors as discussed in Section~\ref{sect:science_far_detector} and stated in Table~\ref{tab:syst} are considered. For the neutrino and antineutrino beam, separate nuisance parameters for the three normalisations (overall normalisation, NC normalisation and $\nu_\tau$~CC normalisation) and the neutrino flavour identification performance (ParticleID skew) are used.
Normal neutrino mass ordering is assumed for all presented $\delta_{\textrm{CP}}$ sensitivity figures.

\begin{figure}
  \centering
  \includegraphics[width=11cm,trim={0.0cm 0.0cm 1.5cm 1.0cm},clip]{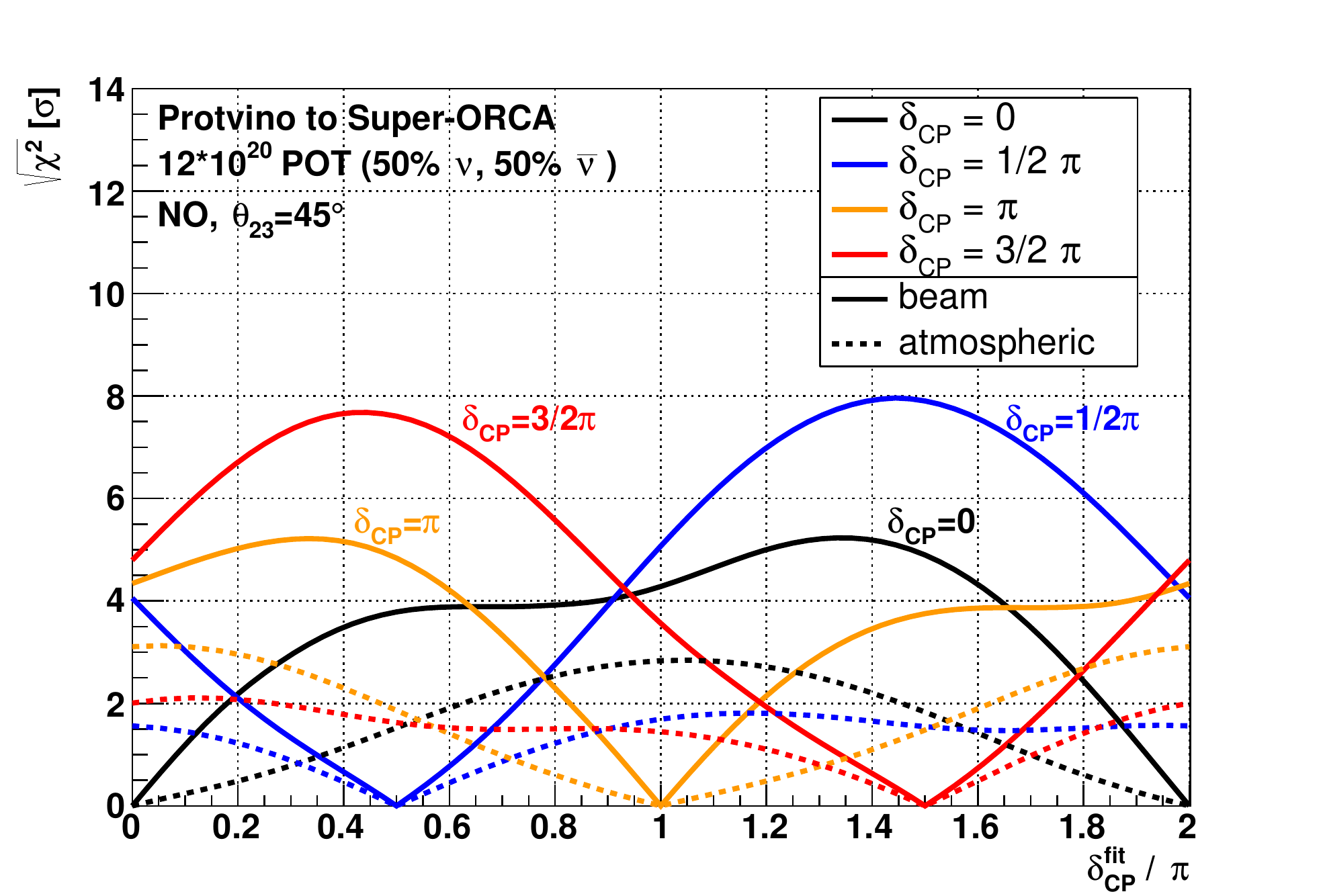}
  \caption{Sensitivity to exclude certain values of $\delta_{\textrm{CP}}$ with Super-ORCA after 3 years
of data taking as a function of the tested $\delta_{\textrm{CP}}^{\rm fit}$ for 4 example values of the true $\delta_{\textrm{CP}}$.
A 450\,kW beam from Protvino with 1.5 years in neutrino mode and 1.5 years in antineutrino mode is assumed (solid lines). 
The corresponding sensitivity for Super-ORCA using 3 years of atmospheric neutrino data is shown by dashed lines for comparison.
          }
  \label{fig:superorca_sensitivity_1}
\end{figure}

Figure~\ref{fig:superorca_sensitivity_1} shows the expected sensitivity to distinguish between different $\delta_{\textrm{CP}}$ values with Super-ORCA after 3 years of data using the Protvino neutrino beam. The largest sensitivity is achieved between $\delta_{\textrm{CP}}=90^\circ$ and $\delta_{\textrm{CP}}=270^\circ$, which correspond to the smallest and largest oscillation probabilities at the first oscillation maximum.
For comparison, also the $\delta_{\textrm{CP}}$ sensitivity for Super-ORCA using atmospheric neutrinos is shown. The operation of Super-ORCA with the Protvino neutrino beam significantly improves the $\delta_{\textrm{CP}}$ sensitivity compared to the measurement with atmospheric neutrinos due to the ability to control the beam polarity ($\nu$ and $\bar{\nu}$ modes).

\begin{figure}
  \centering
  \includegraphics[width=11cm,trim={0.0cm 0.0cm 1.5cm 1.0cm},clip]{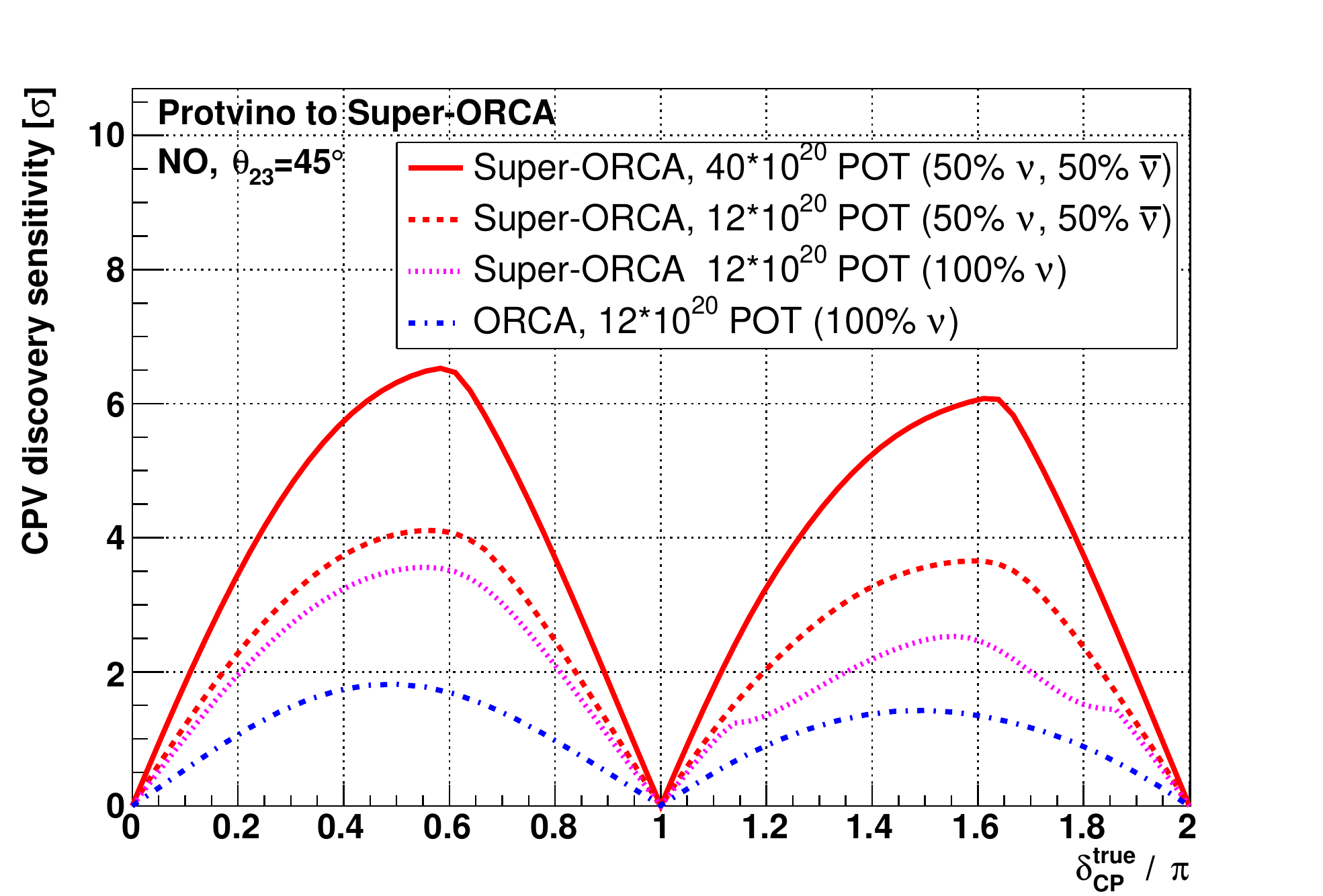}
  \caption{Sensitivity to detect CP violation by operating ORCA 3 years (100\% $\nu$ beam, shown by dash-dotted line), Super-ORCA 3 years (dotted line for 100\% $\nu$ beam and dashed line for 50\% $\nu$/50\% ${\bar \nu}$ beam) and Super-ORCA 10 years (solid line for 50\% $\nu$/50\% ${\bar \nu}$ beam) in a 450\,kW beam from Protvino. 
  The sensitivity is shown as a function of the true $\delta_{\textrm{CP}}$ value ($\delta_{\textrm{CP}}=0$ and $\delta_{\textrm{CP}}=\pi$ correspond to CP conservation).
	   }
  \label{fig:superorca_sensitivity_2}
\end{figure}

The sensitivity to discover CP violation is shown in Figure~\ref{fig:superorca_sensitivity_2} for 3 and 10 years of operation with an equal share between neutrino and antineutrino beam from Protvino. For comparison, also the sensitivity for running only in neutrino mode without antineutrino mode is shown. The kinks (magenta dashed curve, $\delta_{\textrm{CP}}^{true}/\pi \approx 1.13$ and $1.87$) are caused by the $\delta_{\textrm{CP}}-\theta_{13}-\theta_{23}$ degeneracy,
which is resolved when combining neutrino and antineutrino data.

\begin{figure}
  \centering
  \includegraphics[width=11cm,trim={0.0cm 0.0cm 1.5cm 1.0cm},clip]{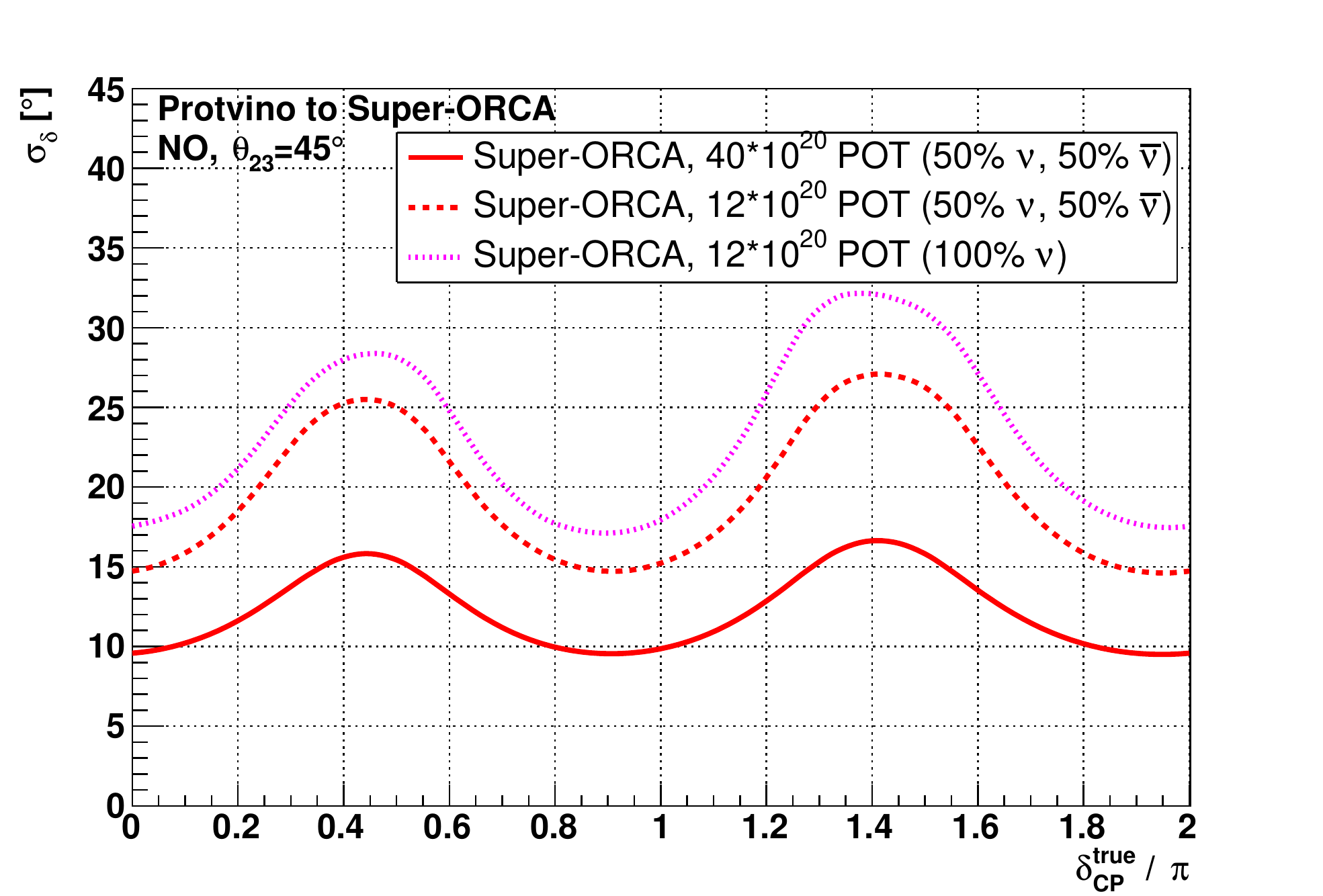}
  \caption{Resolution on $\delta_{\textrm{CP}}$ as function of the true $\delta_{\textrm{CP}}$ value
   for Super-ORCA with the 450\,kW beam operating for 3 years with 100\% $\nu$ beam (dotted line) and 50\% $\nu$/50\% ${\bar \nu}$ beam (dashed line) and 10 years with 50\% $\nu$/50\% ${\bar \nu}$ beam (solid line).
	   }
  \label{fig:superorca_cp_resolution}
\end{figure}

The expected $\delta_{\textrm{CP}}$ resolution reached after 3 and 10 years of running is shown in Figure~\ref{fig:superorca_cp_resolution}. The best measurement precision is achieved for $\delta_{\textrm{CP}}=0^\circ$ and $\delta_{\textrm{CP}}=180^\circ$ with a resolution of $\sigma_{\delta_{\textrm{CP}}} \approx 10^\circ$ after 10 years, while for $\delta_{\textrm{CP}}=90^\circ$ and $\delta_{\textrm{CP}}=270^\circ$ a resolution of $\sigma_{\delta_{\textrm{CP}}} \approx 16^\circ$ is achieved. 
The systematics limiting the $\delta_{\textrm{CP}}$ resolution are the uncertainty on $\theta_{13}$ (mainly for $\delta_{\textrm{CP}}=0^\circ$ and $\delta_{\textrm{CP}}=180^\circ$), the $e$/$\mu$ energy scale skew (mainly for $\delta_{\textrm{CP}}=90^\circ$ and $\delta_{\textrm{CP}}=270^\circ$) and the true value of $\theta_{23}$ (full $\delta_{\textrm{CP}}$ range). The $\delta_{\textrm{CP}}$ resolution is about $1^\circ$ better for $\theta_{23}$ in the first octant compared to $\theta_{23}=45^\circ$ or $\theta_{23}$ in the second octant.
The impact on the $\delta_{\textrm{CP}}$ resolution due to a larger uncertainty on the tau normalisation uncertainty of 20\% instead of 10\% is found at maximum to be $0.5^\circ$ (for $\delta_{\textrm{CP}}=0^\circ$ and  $\delta_{\textrm{CP}}=180^\circ$) after 10 years of operation.

\section{Summary}
\label{sect:summary}
The Protvino accelerator facility is well suited for conducting experiments with GeV neutrino beams and has a strong potential to make important contributions to modern neutrino physics, competing with facilities such as Fermilab and J-PARC.
The distance from Protvino to the ORCA neutrino detector in the Mediterranean Sea is 2595 km,
which is ideal for a long-baseline neutrino experiment employing ORCA as a far detector.
Such an experiment promises an outstanding sensitivity to neutrino mass ordering,
easily reaching a 5$\sigma$ significance level even with a relatively low intensity beam (90 kW).
With a sufficiently long beam exposure ($\approx$ 4 yr $\times$ 450 kW),
a 2$\sigma$ sensitivity to leptonic CP violation ($\delta_{\textrm{CP}}$) can also be reached,
which is comparable with the projected sensitivity of the T2K and NO$\nu$A experiments.
Unique characteristic features of P2O include 1) the longest baseline; 2) the highest energy of the oscillation maximum; and 3) the highest neutrino event statistics due to the large far detector installed in the open sea.

A new neutrino beamline will need to be constructed at Protvino in order to produce a neutrino beam focused in the direction of ORCA.
Achieving a competitive sensitivity to CP violation will require an increase of the accelerator beam power from 15 kW (current value) up to at least 90 kW.
Such an upgrade appears technically feasible.
With a 90 kW beam, ORCA will detect $\sim$ 4000 beam neutrino events per year,
of which about 700 are electron neutrinos
(for the case of normal mass ordering, positive beam polarity).
A near detector is proposed to be constructed a few hundred meters downstream from the proton target in order to monitor the initial parameters of the P2O neutrino beam, study neutrino interactions with matter, and perform other measurements with the neutrino beam, including sterile neutrino searches.

The sensitivity of P2O to $\delta_{\textrm{CP}}$ could be further enhanced by means of an upgrade of the ORCA detector.
Preliminary studies suggest that a 6$\sigma$ sensitivity to CP violation and a 10$^\circ$--17$^\circ$ resolution on $\delta_{\textrm{CP}}$ could be reached
using a 10 times denser version of ORCA with a fiducial volume of 4 Mt after 10 years of operation with a 450 kW beam.
This is competitive with the projected sensitivity of the future experiments DUNE and T2HK.
Similarly to DUNE, T2K/T2HK and ESS$\nu$SB \cite{ESSnuSB},
the best accuracy on $\delta_{\textrm{CP}}$ would be achieved for $\delta_{\textrm{CP}}=0^\circ$ and $180^\circ$.

The sensitivity estimates given here are preliminary and can potentially be improved by optimizing the beamline design and the data analysis pipeline.
Such potential improvements will be explored in a forthcoming study.
The possibility of a non-zero off-axis angle will also be studied.

This letter of interest emphasizes the synergistic potential of the existing accelerator and detector infrastructure: the U-70 proton synchrotron at Protvino and the KM3NeT/ORCA detector in the Mediterranean Sea.
Thanks to the large instrumented volume of ORCA (8~Mt), the beam intensity required for the P2O experiment is relatively small compared to that required for 50 kt scale experiments such as T2K and DUNE.
This allows to re-use most of the existing accelerator infrastructure at Protvino.
In this regard, the construction of such a neutrino beamline at Protvino appears as a good cost-efficient strategy to maximize the scientific output of the Protvino accelerator complex as well as that of ORCA.



\end{document}